\documentclass{article} 

\usepackage{booktabs}
\usepackage[T1]{fontenc}
\usepackage{graphicx}
\usepackage{subfig}
\usepackage{hyperref}
\usepackage{oxford2} 
\usepackage{threeparttable} 
\newcommand*{\thead}[1]{\multicolumn{1}{l}{\bfseries #1}} 
\usepackage{bm} 
\usepackage{mathtools} 
\usepackage{amssymb} 

\usepackage[symbol]{footmisc} 

\usepackage[affil-it]{authblk} 

\usepackage{titlesec}

\titleformat{\paragraph}
{\normalfont\normalsize\bfseries}{\theparagraph}{1em}{}
\titlespacing*{\paragraph}
{0pt}{3.25ex plus 1ex minus .2ex}{1.5ex plus .2ex}

\usepackage[a4paper]{geometry} 

\usepackage{siunitx}
\usepackage{verbatim} 

\title{A Data-Driven Supply-Side Approach for Measuring Cross-Border Internet Purchases}

\author[1,2,6]{Q.A. Meertens\thanks{E-mail: \texttt{q.a.meertens@uva.nl}; Corresponding author}}
\affil[1]{Center for Nonlinear Dynamics, Economics and Finance, University of Amsterdam}
\affil[2]{Department of Economics, Business Statistics and National Accounts, Statistics Netherlands, The Hague}

\author[1,3,4]{C.G.H. Diks}
\affil[3]{UvA Institute for Advanced Study, University of Amsterdam}
\affil[4]{Tinbergen Institute, Amsterdam}

\author[5, 6]{H.J. van den Herik}
\affil[5]{Leiden Centre of Data Science, Leiden University}

\author[6, 7]{F.W. Takes}
\affil[6]{Leiden Institute of Advanced Computer Science, Leiden University}
\affil[7]{CORPNET, University of Amsterdam}

\begin{document}

\maketitle

\section*{Abstract}
The digital economy is a highly relevant item on the European Union's policy agenda. Cross-border internet purchases are part of the digital economy, but their total value can currently not be accurately measured or estimated. Traditional approaches based on consumer surveys or business surveys are shown to be inadequate for this purpose, due to language bias and sampling issues, respectively. We address both problems by proposing a novel approach based on supply-side data, namely tax returns. The proposed data-driven record-linkage techniques and machine learning algorithms utilize two additional open data sources: European business registers and internet data. Our main finding is that the value of total cross-border internet purchases within the European Union by Dutch consumers was over EUR~1.3 billion in 2016. This is more than 6 times as high as current estimates. Our finding motivates the implementation of the proposed methodology in other EU member states. Ultimately, it could lead to more accurate estimates of cross-border internet purchases within the entire European Union.\footnote[3]{The views expressed in this paper are those of the authors and do not necessarily reflect the policy of Statistics Netherlands.}

\vspace{10pt}
\noindent \emph{Keywords: cross-border e-commerce, data-driven record linkage, digital economy, machine learning, official descriptive statistics, online consumption}

\newpage




\section{Introduction}\label{sec_introduction}

Accurate estimates of cross-border online consumption within the European Union (EU) are difficult to obtain. Recently, accurately estimating cross-border online consumption has become more important for two reasons. The first reason is that consumption through online channels of both goods and services is increasing within the EU, especially across borders. Consumers have more access to the internet, shipping costs are decreasing and payment services converge across countries \cite{marcus2016commerce,cardona2016delivery,martikainen2015convergence}. Accurate estimates of cross-border online consumption are thus of increasing importance for adequately reporting on national accounts by national statistical institutes. The second reason is that cross-border online trade is nowadays a highly relevant item on the EU's policy agenda when considering the EU Digital Single Market (European Commission, COM/2010/0245). Therefore, getting grip on cross-border online consumption through reliable estimates is essential in quantifying the effect of any new policy. The need for accurate estimates on aspects of the digital economy within the European Union is emphasized by the \citetext{eu2015monitoring}.

Traditional methods for estimating consumption are based on either consumer surveys or business surveys. One EU-wide consumer survey on cross-border online consumption is conducted by Ecommerce Europe (\url{https://www.ecommerce-europe.eu}). In the Netherlands, this survey is conducted by market research institute GfK (\url{https://www.gfk.com}). It is commissioned by the national e-commerce association in the Netherlands, Thuiswinkel.org (\url{https://www.thuiswinkel.org}), on behalf of Ecommerce Europe. The estimates of total cross-border online consumption are based on asking consumers how much they spent at foreign webshops over a fixed time period in the past. However, webshops selling goods or services typically operate in a country using a website in the regional language \cite{schu2017foreign}. This observation corresponds to the main impediment of online consumption, which is foreign language, rather than security reasons, shipping costs, geographical distance or available payment services \cite{gomez2014drivers}. Therefore, it is difficult for a consumer to distinguish between domestic and foreign webshops, as both will be presented in the same, regional language. A consumer-survey approach thus leads to a severe downward bias in measuring cross-border online consumption \cite{unctad2016ecommerce}.

A solution suggested by \citetext{unctad2016ecommerce} is to use business surveys instead. For measuring cross-border online consumption of consumers in a single country, large companies within the EU should report their sales to consumers per EU member state. This places a huge burden on companies. Moreover, the approach poses significant challenges on any correction for sampling probabilities and biases if, for example, the population of the existing ICT business survey would be used. The referenced population is the result of sampling and stratification with respect to (a) economic activity and (b) either relative turnover or number of employees. The stratified sampling probabilities with respect to size in one country have to be transformed into that of the total online sales in another country. Given large differences between countries in this regard, it seems infeasible to arrive at accurate estimations for cross-border online consumption for each EU member state using the suggested solution. It thus renders traditional official statistical methods inadequate.

Based on the shortcomings of traditional methods, an adequate method to measure cross-border online consumption should at least meet the following three requirements. First, the method should be based on supply-side data, to prevent a strong downward bias due to a webshop's availability in different languages. Second, it should be based on existing data that were collected to accurately measure the sales of companies across borders. A reliable administrative or other integral data source would be preferable, to prevent dealing with sampling issues.  Third, the data would have to be available to national statistical institutes across the European Union in order to longitudinally monitor cross-border online consumption.

Motivated by these three requirements, we propose to use supply-side data in the form of tax returns filed by foreign companies. The EU system of value added tax (VAT) states that any company established in the EU that is involved with cross-border intra-community supplies to consumers has to pay VAT in the country of destination, through filing a tax return (European Commission, Council Directive 2006/112/EC). The threshold value on total turnover from sales to consumers, above which filing a tax return is mandatory, is either EUR~35,000 or EUR~100,000, depending on the country of destination. For foreign companies selling to consumers in the Netherlands, the threshold value on total sales in the Netherlands equals EUR~100,000. The Dutch Tax and Customs Administration collects such tax returns, which are then made available to Statistics Netherlands. Here we note that using these data restricts us to measuring cross-border \emph{online consumption of goods} (henceforth referred to as cross-border internet purchases). The approach that we will propose cannot be applied to measure cross-border \emph{online consumption of services}.

The main challenge in using filed tax returns is to identify webshops. For this identification, we propose a two-phase approach. In the first phase, the aim is to select the companies that are economically active in retail trade, according to the NACE (Rev. 2) classification of industries. 
NACE is the acronym for \emph{Nomenclature statistique des Activit\'es \'economiques dans la Communaut\'e Europ\'eenne} and is thus the "statistical classification of economic activities in the European Community". The filed tax returns do contain a classification of economic activity for each company. However, it is based on a statistical classification of industries as of 1974. As webshops did not exist in 1974, the classification is inadequate in identifying webshops. We propose a data fusion approach, merging the companies filing tax returns with a complete list of European companies active in retail trade. The list can be obtained from any business register (BR) containing the company names, country of establishment and the NACE classification of industries for every company in the EU. We choose to use the global ORBIS database (\url{http://bvdinfo.com/orbis}), maintained by Bureau van Dijk, as it is an open data source. We remark that the legal name of a company can be registered differently in the BR compared to the filed tax returns. As a result, merging legal company names from the two datasets is not straightforward due to challenges related to record linkage. Mostly, only three types of differences occur. The first type of difference results from different uses of non-alphanumeric characters, such as a period ($.$), a hyphen ($-$) or an ampersand (\&). The second type of difference is the notation (abbreviated or fully written out) of the type of business entity, at the end of a legal company name. An example is \emph{ltd} instead of \emph{limited}. The third type of difference is spelling, in particular for non-ASCII characters such as the German \emph{\"u}, which occurs either as \emph{ue} or simply \emph{u}. Other common differences in spelling are the use or omission of white spaces, such as writing \emph{web shop} instead of \emph{webshop}. Other types of occurring differences, such as an updated company name which has been processed in only one of the two databases, are more complicated. We use text mining and data-driven record-linkage techniques to neutralize the first three types of differences. A significant part of the methods section describes this first phase.

A second phase is necessary, because being a foreign company active in retail trade, exporting to the Netherlands is not equivalent to being a foreign webshop, exporting to the Netherlands. We might assume that a foreign webshop that sells goods to consumers in the Netherlands will file a tax return (if the annual turnover is above the threshold value) and will be registered as a retail company in the BR, either as principal or secondary economic activity. However, a foreign company that files a tax return in the Netherlands and that is registered as a retail company in the BR, might not necessarily be active as an online retailer \emph{in the Netherlands}. As a result, merging tax returns and the list of retail companies in the BR might lead to false positives. Therefore, we propose a second phase based on internet data. This  phase roughly consists of two steps. The first step is to find the web page of a company based on the legal company name. The second step is to assess whether the web page belongs to a webshop, using text-based features obtained from the HTML-code of the web page.

In the implementation of the two phases described above, fine tuning of parameters is still required. We provide for phase an example. In the first phase, for example, it has to be specified how similar two legal company names should be in order to be considered identical. It requires choosing an optimal threshold value for the similarity score of two legal company names. In the second phase, for example, a function has to be chosen that maps text-based features obtained from the data to a binary categorization, namely webshop or not a webshop. In both phases, the fine tuning of parameters is achieved by using machine learning techniques. The challenge in using machine learning is choosing a suitable algorithm from the wide variety of machine learning algorithms that exist. A solution is to compare the goodness of fit of a variety of machine learning algorithms. The algorithm that performs best according to the goodness of fit on the data is chosen to fine tune the parameters.

The proposed approach is a combination of using supply-side data measuring internet sales and data-driven methods to identify foreign webshops. We claim that the proposed approach yields more accurate estimates of cross-border internet purchases than existing or earlier suggested approaches. It should be noted that the proposed approach might still result in an estimate of cross-border internet purchases with a small downward bias, for which two reasons can be identified. First, companies with sales below the threshold value in the country of destination do not have to file a tax return. The internet purchases at such small companies are therefore missing in a measurement based on filed tax returns. We remark that there may exist many companies of this type. Second, the reported turnover from sales to consumers might be inaccurate, potentially leading to an underestimation of the total cross-border internet purchases. The second effect, however, is expected to be minimal, due to strict law enforcement by and collaboration between tax authorities in the EU. Whatever the cause, we do not aim to correct for any of these two biases, as there is no data available to estimate them. Moreover, our main aim is to demonstrate the severe downward bias of consumer-survey approaches compared to a supply-side approach in estimating cross-border internet purchases within the EU. We therefore do not mind if our supply-side approach still yields a conservative estimate.

The remainder of the paper is organized as follows. In Section \ref{sec_data}, we describe the three supply-side datasets that have been used. We also describe how the training and test datasets needed for the machine learning algorithms are obtained. In Section \ref{sec_methods}, the data-driven methods to identify foreign webshops are thoroughly discussed. In Section \ref{sec_results}, we present the results of applying the proposed approach to the economy of the Netherlands. In addition, the results are compared with results from an existing consumer-based approach to measure cross-border internet purchases in the Netherlands. Section \ref{sec_conclusion} concludes by encouraging the implementation of our method in other countries within the EU and discussing possible further research.




\section{Data}\label{sec_data}

In this section, we will describe the three supply-side datasets that are used to measure cross-border internet purchases. In Subsection \ref{sec_data_description}, the turnover distribution is depicted and some summary statistics are mentioned. In Subsection \ref{sec_data_train}, we explain how the training and validation dataset were constructed. In Subsection \ref{sec_data_test}, we describe the test dataset.

\subsection{Data Description}\label{sec_data_description}
\begin{figure}[t]
\centering
\subfloat{\includegraphics[width=0.32\textwidth]{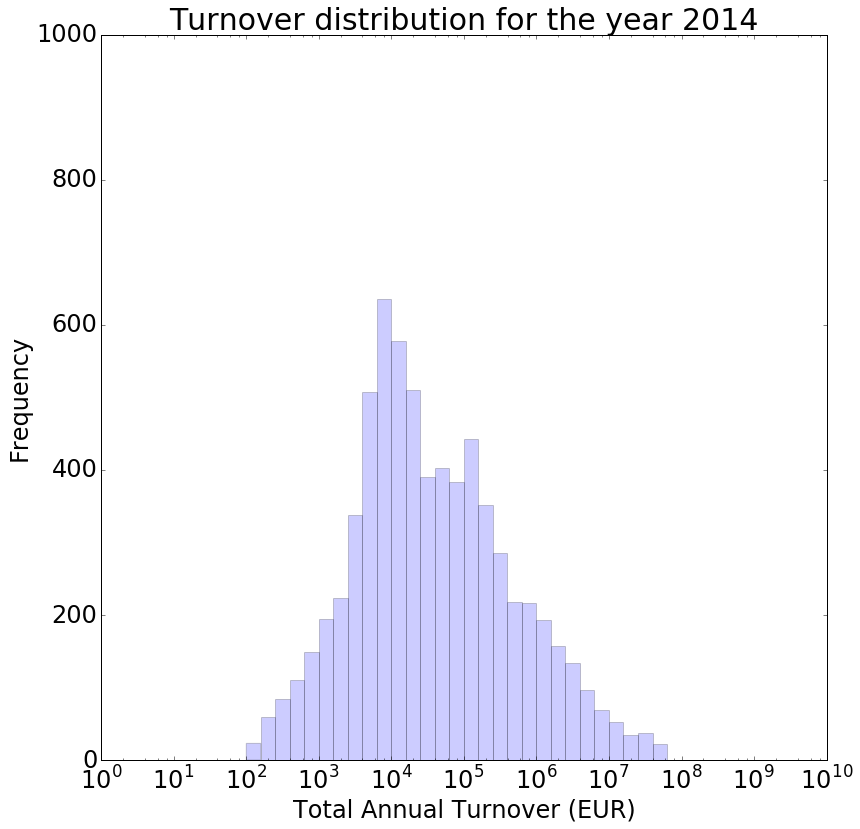}}
\subfloat{\includegraphics[width=0.32\textwidth]{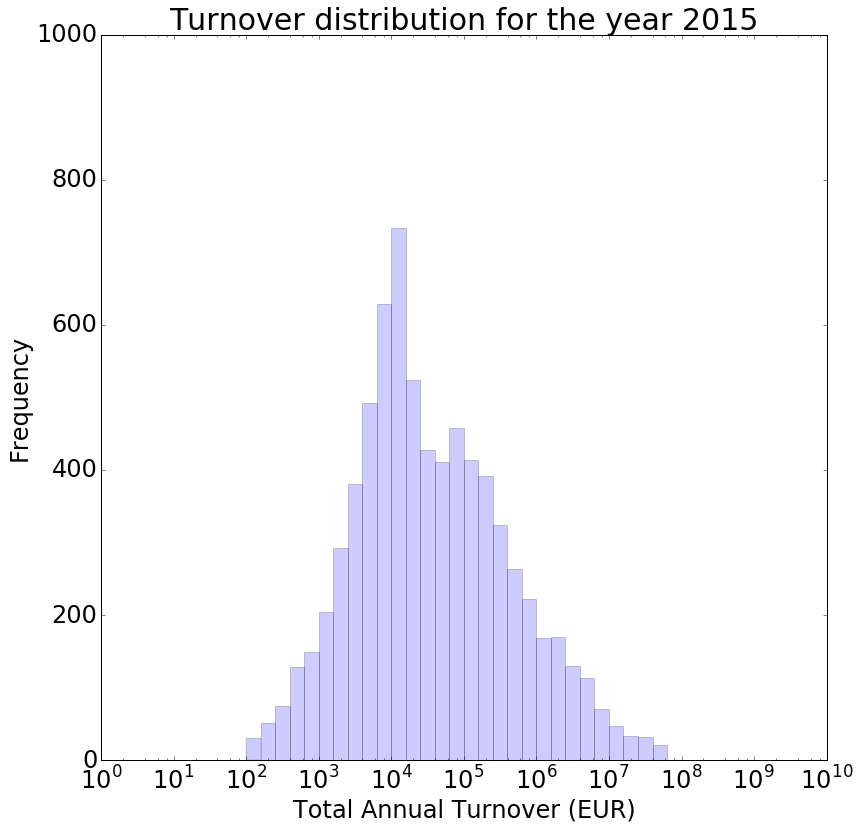}}
\subfloat{\includegraphics[width=0.32\textwidth]{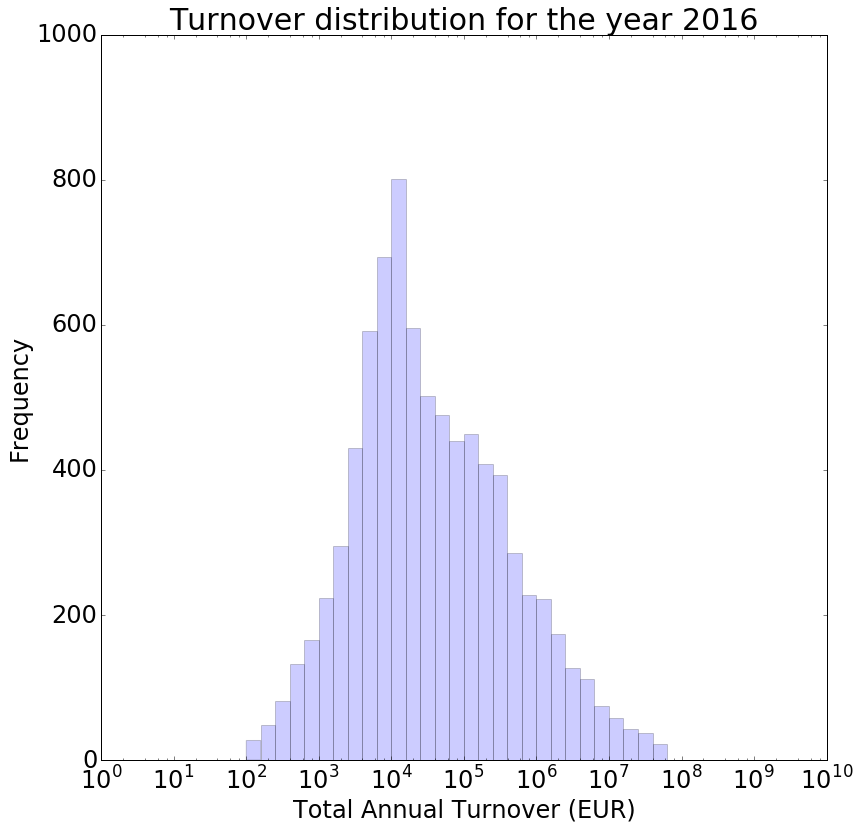}}
\caption{The distributions of annual turnover of foreign companies that filed a tax return in the Netherlands for 2014, 2015 and 2016. Note that the horizontal axis is a logarithmic scale. Companies that reported negative or zero turnover are not shown. Due to privacy legislation, bins containing less than 20 companies are removed.}
\label{fig_data_distr}
\end{figure}

The data used to measure cross-border internet purchases are tax returns filed in the Netherlands by foreign companies established in the EU. This dataset of tax returns contains legal company names and the annual turnover from sales in the Netherlands of goods taxed at low or high tariff. The data are extracted from tax returns filed for 2014, 2015 and 2016. The dataset contains 197,424 filed tax returns from $22,440$ unique companies. Logically, this dataset of tax returns filed in the Netherlands is not openly available, due to strict privacy legislation. Under severe restrictions (among others, anonymizing) and obeying serious impositions (such as suppressing extreme values), we are permitted to present aggregated figures on the data. When relevant, we reveal the criterion for which we suppressed information. Hence, in Fig. \ref{fig_data_distr} we show the distribution of the annual turnover for each of the years 2014, 2015 and 2016. Furthermore, Table \ref{tab_data_distr} displays summary statistics of the dataset of tax returns filed in the Netherlands.

\begin{table}[t]
\centering
\caption{Summary statistics (mean, median and 10th and 90th percentile of annual turnover) of the tax returns filed in the Netherlands by foreign companies in 2014, 2015 or 2016 are displayed. The set of companies filing a tax return in a certain year in denoted by $I$. The subscript denotes whether the annual turnover in the given period is $0$, lower than $0$, or larger than $0$. As usual, $|X|$ is used to denote the number of elements in a set $X$.}
\label{tab_data_distr}
\begin{threeparttable}
\begin{tabular}{lrrrrrrrrrr}
\hline
\thead{Year} & $\bm{|I|}$ & $\bm{|I_0|}$ & $\bm{|I_{<0}|}$ & $\bm{|I_{>0}|}$ & \thead{Mean} & \thead{Median} & \thead{10th perc} & \thead{90th perc} \\
\hline
2014 & 16,023 & 8,969 & 86 & 6,968 & 1,904,860 & 25,987 & 1,755 & 1,365,217 \\
2015 & 17,313 & 9,771 & 80 & 7,462 & 1,603,908 & 25,166 & 1,783 & 1,218,926 \\
2016 & 18,939 & 10,626 & 104 & 8,209 & 1,488,351 & 24,790 & 1,879 & 1,143,156 \\
\hline
\end{tabular}
\end{threeparttable}
\end{table}

In addition, two open data sources have been used. The first open data source is the set of home pages of web sites of foreign companies that filed at least one tax return in the Netherlands for 2014, 2015 or 2016. The pages were downloaded on April 19, 2017.

The second open data source is ORBIS, a global corporate database maintained by Bureau van Dijk (\url{http://bvdinfo.com/orbis}).
It contains detailed corporate information on over 200 million private companies worldwide. The database has been claimed to ``suffer from some structural biases'' \cite{ribeiro2010oecdorbis}. However, for European companies having more than EUR 100,000 annual turnover, the dataset is close to being complete \cite{garcia2016coverage}. Data on smaller foreign companies are not needed in our analysis, as they do not have to file tax returns in the Netherlands. The ORBIS database is used, because it contains the principal and secondary NACE (Rev. 2) codes for companies established in the EU. The NACE code can be used to select all active and inactive companies established in the EU that are principally or secondarily economically active in retail trade. The result is a dataset of $6,996,468$ companies, in which companies established in the Netherlands are excluded. This dataset, including each company's country of establishment, was extracted from ORBIS on June 24, 2017.

For our purposes, any business register (BR) containing the company names and country of establishment of every retail company in the EU would suffice, as long as the retail companies (according to NACE Rev. 2) can be identified as such. Therefore, we will henceforth refer to the ORBIS data as ``the BR''.

\subsection{Training and Validation Dataset}\label{sec_data_train}
In order to train classification algorithms, a labeled dataset is required. Since no such dataset existed, it was manually constructed as follows. The filed tax returns contain a classification of economic activity according to the (outdated) statistical classification of industries as of 1974. At first glance, most webshops seemed to be classified as \emph{Retail Trade}, many as \emph{Wholesale trade} and some as another type of industry according to the outdated classification. We constructed a training dataset of 180 companies by manually categorizing all companies of which the total annual turnover in at least one year exceeded an industry-dependent threshold value (see Table \ref{tab_data_train}, column 2). Table \ref{tab_data_train} (column 3) displays the number of companies manually categorized per type of industry. In fact, two manual categorizations were made for each company included in the training dataset as presented in Table \ref{tab_data_train}. The first categorization is whether the company is economically active as a retail company according to the BR. We remark that the economic activity reported in the BR might be different from the one found in the tax return dataset. The second categorization is whether the company is actually a webshop or not, based on manually searching the internet.

Within the training dataset, 76 webshops were identified. Their total turnover in 2016 was equal to EUR 724,542,550. The validation dataset will be obtained from the training dataset when applying stratified $5$-fold cross-validation. It is described in more detail in Section \ref{sec_methods_model}.

\begin{table}[t]
\centering
\caption{The number of companies per industry class (using a classification from 1974) included in the training dataset. The threshold value of the annual turnover is used as a selection criterion for a company to be included in the training dataset.}
\label{tab_data_train}
\begin{threeparttable}
\begin{tabular}{lcrrr}
\hline
\thead{Industry (1974)} & \thead{Threshold} & \thead{Count} \\
\hline
Retail Trade & EUR \hspace{1pt} 1 million & 100\\
Wholesale Trade & EUR 20 million & 30 \\
Other & EUR 50 million & 50 \\
\hline
Total & -- & 180 \\
\hline
\end{tabular}
\end{threeparttable}
\end{table}

\subsection{Test Dataset}\label{sec_data_test}
To assess the goodness of fit of a classification algorithm, an additional labeled `test' dataset is required. We constructed a test dataset by manually categorizing a set of 80 randomly selected companies not occurring in the training dataset. (One duplicate retail company had to be removed resulting in 79 companies in the test dataset.) During the sampling, the relative frequency of the type of industry in the entire dataset was taken into account. Table \ref{tab_data_test} shows the frequencies and the total number of webshops found among the 79 companies, namely 13.

\begin{table}[t]
\centering
\caption{The number of companies per industry class (using a classification from 1974) included in the test dataset. The frequency of each industry class in the entire dataset of filed tax returns is included, as well as the number of identified webshops per industry class in the test dataset.}
\label{tab_data_test}
\begin{threeparttable}
\begin{tabular}{lcrrr}
\hline
\thead{Industry (1974)} & \thead{Total Frequency} & \thead{Count} & \thead{Webshop Count}\\
\hline
Retail & 1,393 & 19 & 6 \\
Wholesale & 3,329 & 20 & 1 \\
Other & 17,718 & 40 & 6 \\
\hline
Total & 22,440 & 79 & 13 \\
\hline
\end{tabular}
\end{threeparttable}
\end{table}




\section{Methods}\label{sec_methods}

In this section, we will discuss the data-driven methods we have used to estimate cross-border internet purchases within the EU by Dutch consumers. The methods are described in such a way that they can be generalized to other countries within the EU. We present the methods in three parts. In Section \ref{sec_methods_model}, the methods used to identify foreign webshops in the dataset of tax returns are specified. This first section, called \emph{Estimating the industry class}, contains the most important methodological contributions of the paper. Section \ref{sec_methods_correction}, titled \emph{Bias correction}, outlines how we have corrected for biases introduced by inaccuracies in the methods used to identify foreign webshops. In Section \ref{sec_methods_summary}, we briefly summarize our proposed approach for measuring cross-border internet purchases within the EU. Moreover, we motivate why we speak of a \emph{data-driven supply-side} approach.

\subsection{Estimating the Industry Class}\label{sec_methods_model}
The general setup is as follows. Consider a population of $n$ companies indexed by a set $I$. For a company $i \in I$, the industry class is denoted by $s_i \in \mathcal{H}$. The set $\mathcal{H}$ consists only of two industry classes, namely webshops ($s_i = 1$) and other companies ($s_i = 0$). The total turnover from sales of goods taxed at the low or high tariff as reported in the tax returns in year $t$ by company $i$ is denoted by $y_{i,t}$. In the data we have available, $y_{i,t}$ is given for each $i \in I$ and each $t \in \{2014, 2015, 2016\}$. The goal of the paper is to estimate
\begin{equation}\label{eq_goal}
\sum_{i \in I} I(s_i = 1)y_{i,t},
\end{equation}
for each year $t$. In the above, $I(\cdot)$ denotes the indicator function.

The challenge of estimating expression (\ref{eq_goal}) is that the industry classes $s_i$, $i \in I$, are not observed and thus have to be estimated. We propose to estimate $s_i$ in two different ways and combine the two estimates into a single estimate of $s_i$. The first way to estimate $s_i$ is by business registers (BR). This is the main topic of Subsection \ref{sec_methods_model_br}. The second way to estimate $s_i$ is by web sites. This is discussed in Subsection \ref{sec_methods_model_scraper}. In Subsection \ref{sec_methods_model_final}, we propose how to combine the two estimates of $s_i$ into a single estimate. This final, `combined' estimate of $s_i$ is used to evaluate (\ref{eq_goal}).

\subsubsection{Estimating the Industry Class by Business Registers}\label{sec_methods_model_br}

This subsection describes how to estimate the industry class $s_i$ of a company $i \in I$ by business registers. The estimate will be denoted by $\widehat{s_i}^\text{BR}$. To estimate the industry class by business registers, we assume that a webshop can be identified by evaluating whether it is registered in the business register (BR) as a retail company (according to the NACE (Rev. 2) code) in the country in which the office filing the tax return is established. More specifically, we assume that the sales of goods taxed at the low or high tariff as reported in the tax return by companies registered as a retail company in the BR are precisely the cross-border internet purchases within the EU by the consumers of the EU country under consideration. This assumption is used by checking for each $i \in I$ whether it is registered as a retail company in the BR. If so, $\widehat{s_i}^\text{BR} = 1$, otherwise, $\widehat{s_i}^\text{BR} = 0$. In other words, we aim to find the intersection of (1) the companies in $I$ and (2) the companies in the BR registered as a retail company.

The challenge is that the only available data to find the intersection are the names of the companies in each set. There are four main problems in finding the intersection using company names.
\begin{itemize}
\item[A:] The type of business entity might be registered differently in both datasets. A typical example is $\emph{ltd}$ versus $\emph{limited}$.
\item[B:] The name of a company might be spelled differently in both datasets. A typical example occurs for German names, such as \emph{muller} versus \emph{mueller}.
\item[C:] The datasets we use are relatively large. Finding the intersection based on (slightly differently spelled) names is computationally expensive.
\item[D:] Finding the intersection using (slightly differently spelled) names is an optimization problem: one has to determine a threshold on the permitted number of spelling differences between names belonging to the same company.
\end{itemize}
The following four paragraphs propose solutions for these four problems.

\paragraph*{A. Stemming Company Names}

The first problem we aim to solve is that the type of business entity of the same company might be registered differently in both datasets. The two datasets we refer to are the set $N_T$ of legal company names in filed tax returns and the set $N_R$ of legal company names of EU retail companies in the BR. To be able to compare the type of business entity of two company names, we first have to isolate the type of business entity from the company name. The type of business entity can be isolated by a technique called \emph{stemming}, as the type of business entity is usually denoted at the end of a legal company name. Before stemming is applied, all non-alphanumeric characters in the elements of both $N_T$ and $N_R$ are replaced by white spaces. Then, all leading, trailing and duplicate white spaces are removed. Denote again $N_T$ and $N_R$ by the resultant sets.

Next, stemming is applied on all company names in $N_T$ and $N_R$. Our three-step approach is based on the first published stemming algorithm \cite{lovins1968development} as well as on Porter's stemming algorithm \cite{porter1980algorithm}. The latter still is ``the most common algorithm for stemming English'' \cite[p. 32]{manning2009IR}. In the first step, a country-dependent list of most common end-of-string words, or suffixes, of variable length in the BR is constructed. Only the BR is used, as the dataset of filed tax returns does not contain a company's country of establishment. As the names of types of business entities strongly differ per country, the types of business entities in smaller countries might not show up as most common suffixes. It is assumed that a suffix, if present, consists of at least one and at most four words. The most common suffixes are complemented with a list of known types of business entities per EU member state, obtained from \url{https://en.wikipedia.org/wiki/List_of_business_entities}.
In the second step, the obtained list of suffixes is reduced to a list of most common starts of suffixes. It not only reduces the computation time of removing a suffix, but it also increases the number of suffixes found. This step is inspired by Porter's algorithm for stemming English \cite{porter1980algorithm}. In the third and final step, suffixes are removed from company names in both lists by searching for any of the most common starts of suffixes. It is ensured that the remaining \emph{stems} of company names are never empty.

The above defines a two-valued function $f = (s, t)$, yielding for any name $a \in N_T \cup N_R$ the stem $s(a)$ of $a$ and the start $t(a)$ of the type of business entity of $a$. Define a relation $\sim$ on the image $\text{im}(t)$ of $t$, where $x \sim y$ if $x$ is an abbreviation of one of the types of business entities $y$ is the start of, or vice versa. The relation $\sim$ is reflective and symmetric, but not necessarily transitive. The class $[x]$ of an element $x$ in the image of $t$ is defined as $[x] \coloneqq \{y \in \text{im}(t) : x \sim y\}$. The class $[e(a)]$ is referred to as the suffix class of $a$. Now, the aim is to find for each $a \in N_T$ all $b \in N_R$ for which $s(a)$ and $s(b)$ are very similar and $e(a) \sim e(b)$. That is, determine whether $a$ is similar to any element of $S_R(a) \coloneqq \{s(b) : b \in N_R, e(b) \in [e(a)]\}$. In other words, we are looking for company names having similar stems and identical suffix classes.

\paragraph*{B. Approximate String Matching}

The second problem is that the stems might be spelled differently in both datasets. A typical example would be \emph{webshop ltd} versus \emph{web shop ltd} or \emph{web-shop ltd}. Another typical difference occurs in, for example, German names, such as \emph{muller} versus \emph{mueller}. The variety of possible differences in spelling is huge. The common solution is referred to as approximate string matching.

The idea behind approximate string matching between two sets $A$ and $B$ of strings, is to determine for each element in $A$ the closest element in $B$ according to some string distance metric $d$. An example of such a string distance metric is the Jaccard distance on character $n$-grams, or $n$-shingles \cite[Chapter 3]{leskovec2014mmds}. A \emph{character $n$-gram} is defined as a substring of $n$ consecutive characters in a string. As an example, the set of character $3$-grams, or trigrams, of the string `webshop' is the set \{`web', `eb ', `b s', ` sh', `sho', `hop'\}. For $n \in \mathbb{N}$, write $c_n$ for the functions mapping a string to its set of character $n$-grams. The Jaccard distance on two sets $A$ and $B$ is defined as $d_J(A,B) = 1 - |A \cap B|/|A \cup B|$.  Thus, the Jaccard distance on character $n$-grams $d_{J,n}$ between two strings $a$ and $b$ is defined as $d_{J, n}(a,b) = d_J(c_n(a), c_n(b))$. It can be easily shown that $d_{J, n}$ is a metric for every $n \in \mathbb{N}$. The approximate string match for an element $a \in A$ in the set $B$ according to the metric $d_{J,n}$ would be (one of) the element(s) $b$ in $B$ minimizing $d_{J,n}(a,b), b \in B$. The minimum distance is commonly denoted by $d_{J,n}(a, B)$.

Now, two issues appear. First, the best approximate string match should be categorized as a match or as no match. It means that some threshold value for $d_{J,n}(a,B)$ has to be determined. Second, how to choose $n$, or, more generally, which string distance metric $d$ to choose?

We handle both issues simultaneously by considering multiple string distance metrics and letting a machine learning algorithm determine the optimal threshold values. The string distance metrics considered are (1) normalized Levenshtein (or edit) distance, (2) Jaro-Winkler distance (originally proposed by \citetext{winkler1990metric} and in fact not a metric in the mathematical sense), (3-5) Jaccard distance on sets of character 1-, 2- and 3-grams, and (6-8) cosine distance on term frequency vectors of character 1-, 2- and 3-grams. The Jaro-Winkler, Jaccard and cosine distance are defined as 1 minus the corresponding string similarity measure and always take values in the interval $[0,1]$. The Levenshtein distance is normalized to the interval $[0,1]$ via dividing by the maximum length of the two input strings. All metrics are defined and compared by \citetext{cohen2003comparison} (see also \citeauthor{leskovec2014mmds}, \citeyear{leskovec2014mmds}, pp. 87-93, for a more recent discussion).

At the end of this step, each company in the set of filed tax returns is equipped with an $8$-dimensional vector containing values in the interval $[0,1]$ quantifying the distance (along different metrics) to the set of EU retail companies in the BR. The values will be used as features in the machine learning algorithms as described in Paragraph D of Subsection \ref{sec_methods_model_br}.

Finally, we note that the brute-force approach of computing $d(a, S_R(a))$ as the minimum of $d(a, b)$ for $b \in S_R(a)$ is computationally expensive, as it requires comparing $a$ to each $b \in S_R(a)$. Under the description of problem C below, a more efficient and more elegant approach is described. It is preferred over the brute-force approach.

\paragraph*{C. Min-Hashing and Locality-Sensitive Hashing}

The third problem is the size of the datasets when performing approximate string matching. Each possible combination of a company name from the set of filed tax returns and a company name from the BR has to be assessed by multiple (relatively slow) approximate string matching algorithms. In our case, 22,440 $\times$ 6,996,468 $\approx$ 157,000,000,000 comparisons are needed to evaluate a single approximate string matching algorithm. Depending on the approximate string matching algorithm, this can take up to several days. As we intend to combine the results of different approximate string matching functions, we aim to reduce the computation time.

An efficient and elegant approach is to use locality-sensitive hashing (LSH). The concept of LSH is quite general. In short, it is a form of randomized dimensionality reduction. The approach we have used is outlined in \citetext{bawa2005lsh}. The authors introduce the data structure LSH Forest. It can be queried to retrieve for any object $a$ and any natural number $m$, the $m$ approximately most similar objects in the input dataset according to any metric that induces a locality-sensitive hashing family. One such metric is the Jaccard distance between sets, as introduced above in Paragraph B of Subsection \ref{sec_methods_model_br}.

Below, we briefly describe how the LSH Forest is constructed for the Jaccard distance on character trigrams. We follow the approach outlined in \citetext{bawa2005lsh}, where accompanying details, motivation and additional literature can be found.

Let $a$ be a string containing alphanumerical characters and white spaces only. Enumerate all possible $(26+10+1)^3 = 50,653$ characters trigrams containing only letters, numbers and white spaces. Replace the set of character trigrams $c_3(a)$ by the set of corresponding characters trigram IDs, being the indexes from the enumeration. The randomized dimensionality reduction is to compute for each $c_3(a)$ a $k$-bit min-hash signature. The $k$-bit min-hash signatures are computed as follows. First, randomly choose $k$ hash functions $h_1, \ldots, h_k$ from the family of random linear functions of the form $h(x) = (\alpha x + \beta) \; \text{mod} \; p$, with $a, b$ integers and $p$ a fixed, large prime number. Then, randomly choose $k$ hash functions $g_1, \ldots, g_k$ mapping the values $0, \ldots, p-1$ uniformly at random onto $\{0,1\}$. The $j$-th bit of the $k$-bit min-hash signature of $a$ is then given by $g_j(\min_i(h_j(v_i)))$, where $v$ is the vector containing the IDs of the character trigrams of $a$.

The number $k$ is chosen as small as possible in such a way that every string $s(b)$ for $b \in N_R$ has a unique $k$-bit signature. A maximum of $k = k_m$ is imposed to prevent the signatures from becoming too large. The LSH Tree is defined as the logical prefix tree on all $k$-bit signatures. Note that the leaf nodes correspond to the points $s(b)$, and that the $k$-bit signature of $s(b)$ specifies the path through the tree to the leaf node corresponding to $s(b)$. The LSH Forest consists of $l$ LSH Trees, each constructed with an independently drawn random sequence of hash functions from the described family of hash functions.

Given the stem $s(a)$ of a company name $a \in N_T$, each of the $l$ LSH Trees is updated with an additional leaf node containing (the end point of the path through the LSH Tree specified by the $k$-bit signature of) $s(a)$. The LSH Trees are then simultaneously searched bottom-up starting from the new leaf node, until the $m$ most similar items are identified.

In our analysis, the function MinHashLSHForest from the Python library \emph{datasketch} is used (\url{https://github.com/ekzhu/datasketch}). 
The total number of hash functions is fixed to be 64 and the number of LSH Trees was set to the default value $l=8$. The datasketch implementation then fixes $k = 64/8 = 8$ for the length of the min-hash signatures used to build each of the LSH Trees. The choice of $k = 8$ is relatively small, but works already rather well in our case, as shown in Section \ref{sec_results_br}. The top $m = 100$ most similar leaf nodes $s(b)$ from the LSH Forest are returned for each $s(a)$. Across this set of 100 approximate most similar stems, the closest stems according to each of the 8 string distance metrics as chosen in Paragraph B of Subsection \ref{sec_methods_model_br} are computed.

\paragraph*{D. Machine Learning}

The fourth problem is choosing an optimal threshold of how different two names belonging to the same company are allowed to be. We propose to use machine learning to solve this problem.

More precisely, the aim is to find a classification algorithm $\widehat s_i^{\text{BR}}$ that can accurately predict the industry class $s_i^\text{BR} \in  \mathcal{H}$ using the $8$-dimensional vectors of distances as constructed in Subsection \ref{sec_methods_model_br} so far. The true, unobserved binary value $s_i^\text{BR} \in \mathcal{H}$ captures whether company $i \in I$ is registered in the BR as a retail company. Recall that for each company in both the training set and the test set, the class $\bm{a_i^\text{BR}}$ was observed by manually searching the BR.

To select a classification algorithm and corresponding algorithm parameter settings that are optimal in predicting $s_i^\text{BR}$, the following data-driven approach was chosen. First, ten classification algorithms are chosen to be examined. The ten classification algorithms that we consider are depicted in Table \ref{tab_methods_ml_algs}. We consider the linear classification algorithms Logistic Regression (LR), Linear Discriminant Analysis (LDA) and Linear Support Vector Classification (LinSVC). The nonlinear algorithms implemented are k-Nearest Neighbours (kNN), Multinomial Naive Bayes (MNB), Quadratic Discriminant Analysis (QDA) and Support Vector Classification with Radial Basis Function Kernel (RBFSVC). Further more, we examine three ensemble algorithms, namely Random Forest (RF), Gradient Boosting (GB) and AdaBoost (AB). The algorithms will mostly be referred to using the acronyms between brackets.
The details on the specifications of the classification algorithms can be found in, e.g., \citetext{witten2016datamining}, \citetext{han2011datamining} or \citetext{hastie2009statlearn}. 
We have used the Python library scikit-learn (\url{http://scikit-learn.org/}, version 0.19.1) to implement the ten classification algorithms.

\begin{table}[t]
\centering
\caption{Overview of the ten classification algorithms we have considered, including whether we refer to it as a linear, nonlinear or ensemble algorithm.}
\label{tab_methods_ml_algs}
\begin{threeparttable}
\begin{tabular}{ll}
\hline
\thead{Type} & \thead{Algorithm (Acronym)} \\
\hline
Linear & Logistic Regression (LR) \\
 & Linear Discriminant Analysis (LDA) \\
 & Linear Support Vector Classification (LinSVC) \\
Nonlinear & k-Nearest Neighbours (kNN) \\
 & Multinomial Naive Bayes (MNB) \\
 & Quadratic Discriminant Analysis (QDA) \\
 & Support Vector Classification with Radial Basis Function Kernel (RBFSVC) \\
Ensemble & Random Forest (RF) \\
 & Gradient Boosting (GB) \\
 & AdaBoost (AB) \\
\hline
\end{tabular}
\end{threeparttable}
\end{table}

Then, we specified for each of the ten algorithms a grid of parameter settings to be examined. These grids are depicted in Table \ref{tab_methods_ml_params}. See the scikit-learn documentation for precise parameter specifications. 

Now, for each algorithm and each parameter setting in the parameter grid, stratified 5-fold cross-validation is performed on the training dataset. Cross-validation is used to prevent overfitting. The choice of using 5 folds is based on \citetext{breiman1992kfoldcrossval}, although it might introduce more variance than choosing 10 or 20 folds \cite{kohavi1995stratcrossval}. However, due to the small size of the training dataset, choosing 10 or 20 folds might lead to unstable results. Therefore, we have chosen to use \emph{stratified} 5-fold cross-validation in order to reduce the variance, as suggested by \citetext{kohavi1995stratcrossval}. Furthermore, we optimize parameter settings using mean F1 scores over the 5 folds. We prefer F1 over accuracy due to the low base rate of webshops occurring in the entire dataset. We do not use the common metric AUROC to optimize parameter settings, as it is known to possibly mask poor performance when facing imbalanced data \cite{jeni2013perfmetric}. As our data is in fact strongly imbalanced, due to the low base rate of webshops, it does not seem wise to use AUROC as optimizing metric. Moreover, optimizing AUROC does not, in general, imply optimizing F1 \cite{davis2006f1vsaucroc}. Thus, for each algorithm the parameter setting that maximizes the mean F1 score over the five folds is selected. Subsequently, the mean and standard deviation of F1 scores over the five folds between the ten optimal classification algorithms are compared.

Finally, both the mean F1 score and the standard deviation of F1 scores over the five folds are considered in choosing the final classification algorithm and corresponding parameter settings. If necessary, the local behaviour on the parameter grid is examined to reduce the standard deviation of F1 scores over the five folds. This final classification algorithm is then trained on the entire training dataset. The trained classification algorithm is used to compute the estimate $\widehat{s_i}^\text{BR}$ for each company $i$ not included in the training dataset. Recall that $\widehat{s_i}^\text{BR}$ is an estimate for $s_i^\text{BR}$, which indicates whether company $i$ is registered as a retail company in the BR. In practice, it might be different from the true industry class $s_i$ that we aim to estimate. This is the reason that we will introduce a second estimate $\widehat{s_i}^\text{W}$ in Subsection \ref{sec_methods_model_scraper}. The two estimates $\widehat{s_i}^\text{BR}$ and $\widehat{s_i}^\text{W}$ are combined into a single estimate $\widehat{s_i}^\text{W}$ in Subsection \ref{sec_methods_model_final}.

\begin{table}[t]
\centering
\caption{An overview of the parameter grids for the examined algorithms. In estimating the algorithms LR, LinSVC, RBFSVC, RF and AB, the two class weighing schemes, uniform and balanced, were also included in the parameter grid. See the scikit-learn documentation for parameter specifications.}
\label{tab_methods_ml_params}
\begin{threeparttable}
\begin{tabular}{ll}
\hline
\thead{Algorithm} & \thead{Parameter Grid}\\
\hline
LR & penalty : $\{$l1, l2$\}$;
         $C$: $\{0.001, 0.01, 0.1, 1, 10\}$\\
LDA & \emph{nonparametric}\\
LinSVC & $C$: $\{0.001, 0.01, 0.1, 1, 10\}$\\
kNN & $k$: $\{1, 3, 5, \ldots, 39\}$\\
MNB & $\alpha$: $\left\{10^{-10}, 0.01, 0.1, 1\right\}$\\
QDA & \emph{nonparametric}\\
RBFSVC & $C$: $\{0.01, 0.1, 1, 10, 100\}$;
         $\gamma$: $\{0.001, 0.01, 0.1, 1\}$\\
RF & $n$: $\{50, 100, 200, 500\}$;
         $d$: $\{1, 2, 3, \ldots, 8\}$\\
GB & $n$: $\{50, 100, 200, 500\}$;
         $d$: $\{1, 2, 3, \ldots, 8\}$;
         $\lambda$: $\{0.01, 0.1, 1\}$\\
AB & $n$: $\{50, 100, 200, 500\}$,
         $d$: $\{1, 2, 3, \ldots, 8\}$;
         $\lambda$: $\{0.01, 0.1, 1\}$\\
\hline
\end{tabular}
\end{threeparttable}
\end{table}

\subsubsection{Estimating the Industry Class by Web Sites}\label{sec_methods_model_scraper}

This section describes how to estimate the industry class $s_i$ of a company $i \in I$ by web sites. The estimate will be denoted by $\widehat{s_i}^\text{W}$. To estimate the industry class by web sites, we assume that a webshop can be identified by a shopping cart on the home page, referred to as such in the underlying HTML code. If a shopping cart is found, $\widehat{s_i}^\text{W} = 1$, otherwise, $\widehat{s_i}^\text{W} = 0$.

The main challenge is that filed tax returns do not contain the URL of the web site of a company. We have therefore implemented a method for finding the URL of the web site of a company based on the legal company name. Web scraping is used to automatically look for a shopping cart on the web site. The two tasks are, however, not flawless. Therefore, an additional machine learning approach is used to minimize errors. To summarize, we thus proceed in three steps.
\begin{enumerate}
\item[I.] Find the URL of a company based on the legal company name.
\item[II.] Scrape the HTML code of the URL found in step I and specifically look for a shopping cart.
\item[III.] Use the results from step II in a classification algorithm in order to estimate $s_i$.
\end{enumerate}

The following three paragraphs describe the three steps.

\paragraph*{I. Finding a Company's Web Site}

The first step is to find the URL of the home page of each foreign company filing tax returns. A URL might be available in filed tax returns in some countries. In tax returns filed in the Netherlands, it is not. Therefore, at Statistics Netherlands, software has been developed to find the URL of the home page of a company based on the legal company name. The software, referred to as \emph{URLfinding}, uses Google's Search API. The input is a list of legal company names and the output is a list of URLs per company, ranked according to a score between 0 (definitely not a match) and 1 (definitely a match). The score is referred to as the matching probability, however note that it should not be interpreted as an actual probability in the statistical sense (see ($*$) below). For each company, the URL with the highest matching probability is selected. Both the URL and the corresponding matching probability is returned.

Recalling ($*$) above, we remark that the matching probability is computed using the Random Forest algorithm. The features used are the Jaro-Winkler similarities \cite{winkler1990metric} between the legal company name and some text-based features of the web page, such as the website address, its title and its description.
The training set consists of a few thousand Dutch companies from different industries of different sizes, in terms of number of employees. We explicitly note that the Dutch language of the training set is not necessarily an issue, as discussed in Section \ref{sec_introduction}.
The software has not been developed by the authors and is currently under construction.

\paragraph*{II. Searching for a Shopping Cart}\label{sec_methods_model_scraper_cart}

The second step is to search the identified web site for a shopping cart. For each URL returned by URLfinding the HTML code is downloaded as a raw text file. In the raw text, the occurrences of the words \emph{shop}, \emph{cart} and \emph{basket} in Dutch and English and the word \emph{shopping cart} in German are counted. The full list is \emph{winkel, wagen, mand, shop, cart, bag, basket, warenkorb}. The choice of these three languages is based on the fact that most Dutch citizens mostly speak only Dutch, English and/or German. Note that in modern information retrieval, it is more common to count the occurrences of all words found in a document (cf. chapter 6 in \citetext{manning2009IR}). We have chosen not to follow this approach, as it would lead to serious dimensionality issues: the number of different words (features) would be much larger than the number of documents (companies in the training set).

At the end of this step, each company in the set of filed tax returns is equipped with a vector containing the maximum matching probability and the counts of the eight words mentioned above. The values will be used to train a machine learning algorithms as described in the following paragraph.

\paragraph*{III. Machine Learning}\label{sec_methods_model_scraper_ml}

The third step is to find a classification algorithm $\widehat{s_i}^\text{W}$ that can accurately predict the industry class $s_i \in \mathcal{H}$, using the eight word counts and the maximum matching probability as described in step II above. The true, unobserved industry class $s_i^\text{W} = s_i \in \mathcal{H}$ represents whether company $i \in I$ is a webshop. We recall that for each company in both the training and test set, the class $s_i^\text{W}$ was observed by manually searching the internet.

Before training a classification algorithm, the counts of the words are transformed to real numbers in the interval $[0,1]$ using (normalized) term-frequency $\times$ inverse-document-frequency (TF.IDF) (see \citeauthor{witten2016datamining}, \citeyear{witten2016datamining}, p.314, for a definition). To prevent division by 0 in computing the IDF, a single document containing each of the eight words once is added to the data. The eight TF.IDF values and the maximum matching probability are used as features in fitting classification algorithms on the training data.

The machine learning approach is identical to that described in Paragraph D of Subsection \ref{sec_methods_model_br}. One additional restriction was imposed: if either the maximum matching probability was below $0.5$ or no HTML code could be downloaded, the company was removed from the dataset used to train, test and finally compute $\widehat{s_i}^\text{W}$. For these removed companies, the value of $\widehat{s_i}^\text{W}$ was set to $-1$, to be interpreted as `missing'.

\subsubsection{Constructing the Final Estimate of the Industry Class}\label{sec_methods_model_final}

The two selected classification algorithms, each with the optimal parameter setting, are trained on the entire training dataset. The trained models are used to compute $\widehat{s_i}^\text{BR}$ and $\widehat{s_i}^\text{W}$ on the remaining part of the dataset. Companies whose features, needed for one of the two algorithms, are (partially) missing receive the value $-1$, to be interpreted as `missing', as prediction. It happens for $\widehat{s_i}^\text{BR}$ if the tax-stem of the company has less than three characters. It happens for $\widehat{s_i}^\text{W}$ if the maximum matching probability is below $0.5$ or no HTML code was downloaded. The final, single categorization $\widehat s_i$ is obtained by combining $\widehat{s_i}^\text{BR}$ and $\widehat{s_i}^\text{W}$ as follows:

$$
\widehat s_i \coloneqq
	\begin{dcases*}
    	-1  & when $\widehat{s_i}^\text{BR} = \widehat{s_i}^\text{W} = -1$,\\
    	\widehat{s_i}^\text{BR} & when $\widehat{s_i}^\text{W} = -1$,\\
        \widehat{s_i}^\text{W} & when $\widehat{s_i}^\text{BR} = -1$,\\
        \widehat{s_i}^\text{BR} \land \widehat{s_i}^\text{W} & otherwise.
    \end{dcases*}
$$
The AND-operator $\land$ is computed as the minimum of the two integers. It implies that $\widehat{s_i}$ categorizes a company as a webshop if and only if the company is categorized as such by both $\widehat{s_i}^\text{BR}$ and $\widehat{s_i}^\text{W}$.

Finally, expression \eqref{eq_goal} can be evaluated using $\widehat s_i$ as estimate for $s_i$. This was was the main purpose of Section \ref{sec_methods_model}. A final note is that the manual categorization $s_i^\text{W} = \bm{s_i}$ is used instead of the model $\widehat s_i$ for companies in the training set. We denote by $I_M \subset I$ the set of companies in the training set, being manually categorized. The total (annual) cross-border internet purchases as in expression \eqref{eq_goal} are thus estimated by
\begin{equation}\label{eq_goal_estimate}
\sum_{i \in I_M} I(s_i = 1) y_i + \sum_{i \in I\backslash I_M} I(\widehat s_i = 1) y_i.
\end{equation}
The modelled $\widehat s_i$ introduces a bias in estimating expression \eqref{eq_goal} by expression \eqref{eq_goal_estimate}. The bias is introduced by the second term of expression \eqref{eq_goal_estimate} above. The derivation of the bias and a method to correct for this bias are discussed in Section \ref{sec_methods_correction}.

\subsection{Bias Correction}\label{sec_methods_correction}
As mentioned at the end of Subsection \ref{sec_methods_model_final}, the estimate in expression \eqref{eq_goal_estimate} of total cross-border internet purchases is biased as an estimator for expression \eqref{eq_goal}. Section \ref{sec_methods_correction} has two goals. The first goal is to derive an expression for this bias and propose a method to correct for this bias. The second goal is to estimate the standard deviation of the final estimate of cross-border internet purchases.

For achieving both goals, we follow the approach of \citetext{delden2016errors_published}. The notation used so far is obtained from their paper. We will also introduce the vector notation from their paper. We write $\bm{a_i}$ for the 2-vector $(I(s_i = 1), I(s_i = 0))^T$ and consider the aggregate turnover vector $\bm{y} = \sum_{i \in I} \bm{a_i} y_i$. Similarly, define $\bm{\widehat a_i}$ based on $\widehat s_i$. Expression \eqref{eq_goal_estimate} will thus become the first component of the estimated 2-vector $\bm{\widehat y}$ given by
\begin{equation}\label{eq_hat_y}
\bm{\widehat y} \coloneqq \sum_{i \in I_M} \bm{a_i} y_i + \sum_{i \in I\backslash I_M} \bm{\widehat a_i} y_i.
\end{equation}
In the remainder of Section \ref{sec_methods_correction} only the subset $I\backslash I_M \subset I$ is considered. Hence, any index $i$ will refer to a company not in the training set $I_M$. Consequently, the estimate $\bm{\widehat y}$ will be used to refer only to the second term in the right-hand side of equation \eqref{eq_hat_y}, as the first term does not introduce any bias. Similarly, $\bm{y}$ will be used in the remainder of Section \ref{sec_methods_correction} to refer to $\sum_{i \in I\backslash I_M} \bm{a_i} y_i$.

The approach of \citetext{delden2016errors_published} entails that $s_i$ is considered to be deterministic and $\widehat s_i$ to be stochastic. Consider the classification-error model
$$
P_{ghi} \coloneqq \mathbb{P}(\widehat s_i = h \mid s_i = g), \quad \quad g, h \in \mathcal{H}.
$$
We assume that $P_{ghi}$ does not depend on $i \in I \backslash I_M$. This assumption might be argued to be incorrect for two reasons. First, it is more difficult to find the correct web site for a small company than for a large company. Moreover, a small company that is not a webshop might not even have a web site. Second, the coverage and quality of the BR for smaller companies is significantly lower than for larger companies. Both reasons imply that the probability of a classification error (more specifically, a false negative classification error) increases as turnover decreases. However, estimating $P$ for different turnover classes, as suggested by \citetext{delden2016errors_published}, requires a far larger training dataset, which is quite labour intensive. Moreover, it would also require some large companies \emph{not} to be included in the training dataset. We do not believe that this is wise in our case, as the turnover distribution is highly skewed, as depicted in Fig. \ref{fig_data_distr} in Section \ref{sec_data_description}. Therefore, we conclude that the accuracy of the estimation relies more heavily on accurately classifying larger companies than on accurately estimating $P$.

The resulting 2$\times$2-matrix $P = (P_{gh})_{g,h \in \mathcal{H}}$ is estimated as follows. On the test dataset, $\widehat s_i$ is compared to $s_i$. Denoting by TP, FP, TN, FN the number of true and false positives and true and false negatives, respectively, the estimator $\widehat P$ for $P$ takes the form
$$
\widehat P =
\begin{pmatrix}
\dfrac{\text{TP}}{\text{TP}+\text{FN}} & \dfrac{\text{FN}}{\text{TP}+\text{FN}} \\
\\
\dfrac{\text{FP}}{\text{TN}+\text{FP}} & \dfrac{\text{TN}}{\text{TN}+\text{FP}}
\end{pmatrix}.
$$
For ease of notation, we will write $P$ to refer to the estimated $\widehat P$.

Now, the classification-error model implies $\mathbb{E}(\bm{\widehat a_i}) = P^T\bm{a_i}$ and hence $\mathbb{E}(\bm{\widehat y}) = P^T \bm{y}$. The bias of $\bm{\widehat y}$ as an estimator for $\bm{y}$ equals $\bm{B}(\bm{\widehat y}) = \mathbb{E}(\bm{\widehat y}) - \bm{y} = (P^T - I_2)\bm{y}$, where $I_2$ is the 2$\times$2-identity matrix. This estimator is biased as soon as the classification algorithm makes any mistakes (meaning $P^T \neq I_2$) and $\bm{y}$ is not the eigenvector of $P^T$ corresponding to eigenvalue 1. It implies that in almost all cases, $\bm{\widehat y}$ is indeed a biased estimator for $\bm{y}$ as soon as the classification algorithm makes any mistakes. Denoting by $Q$ the inverse of $P^T$, it follows that $Q\bm{\widehat y}$ is an unbiased estimator for $y$ and $\bm{B}(Q\bm{\widehat y})$ is an unbiased estimator of $\bm{B}(\bm{\widehat y})$.

However, correcting $\bm{\widehat y}$ with $Q$ might increase the variance of the estimator. It might lead to low accuracy in practice. The suggested approach is as follows. Construct bootstrap estimators $\bm{\widehat a_{ir}^*}$, $r = 1, \ldots, R$, by applying the transition matrix $P$ to the predicted $\bm{\widehat a_i}$. It entails that realisations of the alternative classification-error model
given by
$$
\mathbb{P}(\widehat s_{ir}^* = h \mid \widehat s_i = g) = \mathbb{P}(\widehat s_i = h \mid s_i = g) = P_{ghi} = P_{gh}
$$
are considered, where $\widehat s_{ir}^*$ is the industry class corresponding to the 2-vector $\bm{\widehat a_{ir}^*}$. It results in bootstrap replications $\bm{\widehat y_r^*} = \sum{\bm{\widehat a_{ir}^*} y_i}$, $r = 1, \ldots, R$, of the estimated total turnover. The bootstrap bias and variance are given by
$$
\bm{\widehat B_R^*}(\bm{\widehat y}) \coloneqq \bm{B}( \bm{\widehat y^*} \mid \bm{\widehat y}) = m_R(\bm{\widehat y^*}) - \bm{\widehat y}
$$
and
$$
\widehat V_R^*(\bm{\widehat y}) \coloneqq V(\bm{\widehat y^*} \mid \bm{\widehat y}) = \frac{1}{R-1}\sum_{r=1}^R \left(\bm{\widehat y^*} - m_R(\bm{\widehat y^*})\right)\left(\bm{\widehat y^*} - m_R(\bm{\widehat y^*}\right)^T,
$$
where $m_R(\bm{\widehat y^*}) = \sum_r \bm{\widehat y_r^*}/R$ is the sample mean. In the limit of $R \to \infty$, we find
$$
\bm{\widehat B_\infty^*}(\bm{\widehat y}) = \left(P^T - I_2\right) \bm{\widehat y}
$$
and
$$
\widehat V_\infty^*(\bm{\widehat y}) = \text{diag}\left(P^T \bm{\widehat k}\right) - P^T \text{diag}\left(\bm{\widehat k}\right) P,
$$
in which $\bm{\widehat k} = \sum_i \bm{\widehat a_i} y_i^2$. The function $\text{diag}$ maps an $n$-vector $\bm{x}$ to a $n\times n$-matrix with the values of $\bm{x}$ on the diagonal and zeros as off-diagonal elements. Now, in this limit of $R \to \infty$ both bootstrap estimators are biased. The bias of $\widehat V_\infty^*$ as an estimator for the variance of $\bm{\widehat y}$ can be shown to be relatively small and is therefore not corrected \cite[Appendix A4]{delden2015errors_details}. The bias of $\bm{\widehat B_\infty^*}(\bm{\widehat y})$ as an estimator for the bias of $\bm{\widehat y}$ is precisely $P^T$. A second bootstrap estimator $\bm{\widehat b_{ir}^*} \coloneqq Q\bm{\widehat a_{ir}^*}$ is defined, resulting in the bootstrap replications $\bm{\widehat z_r^*} = \sum{\bm{\widehat b_{ir}^*} y_i}$, $r = 1, \ldots, R$ of estimated total turnover. Write $\bm{\widehat B_0} = \bm{\widehat B_R^*}(\widehat y)$ and $\bm{\widehat B_1} = \bm{\widehat B_R^*}(\widehat z)$. The vector $\bm{\widehat B_1}$ is an unbiased estimator of $\bm{B}(\bm{\widehat y})$, but might in practice have a larger variance than the biased estimator $\bm{\widehat B_0}$. To obtain the most accurate results, the goal is to find the optimal value $\lambda = \lambda_\text{opt}$ such that the linear combination $\bm{\widehat B_\lambda} = (1-\lambda)\bm{\widehat B_0} + \lambda \bm{\widehat B_1}$, $\lambda \in [0,1]$, of $\bm{\widehat B_0}$ and $\bm{\widehat B_1}$ minimizes the mean squared error of the first component $(\bm{\widehat B_\lambda})_1$ of $\bm{\widehat B_\lambda}$ as an estimator for the bias $\bm{B}(\bm{\widehat y})$. The mean squared error is given by:
$$
\text{mse}\left((\bm{\widehat B_\lambda})_1\right) = \left\{\bm{B}(\bm{\widehat B_\lambda})\right\}_1^2 + \left\{V(\bm{\widehat B_\lambda})\right\}_{11}.
$$
The following iterative approach is suggested by \citetext{delden2016errors_published}.
\begin{itemize}
\item[0.] Start with $\lambda_\text{old} = 0$.
\item[1.] Compute $\bm{\widehat B} = \bm{\widehat B_0} = \bm{\widehat B_\infty^*}(\bm{\widehat y^*} \mid \bm{\widehat y})$ and $\widehat \Omega = \widehat V_\infty(\bm{\widehat y^*} \mid \bm{\widehat y})$.
\item[2.] Compute $\lambda_\text{new} = \max\{0, \min\{1, (m_1 - m_3 + m_4)/(m_1+m_2-2m_3+m_4)\}\}$, where
\begin{align*}
m_1 &= \left((P^T-I_2)\bm{\widehat B}\right)_1^2,\\
m_2 &= \left((P^T-I_2) \widehat \Omega (P^T-I_2)^T Q^TQ\right)_{11},\\
m_3 &= \left((P^T-I_2) \widehat \Omega (P^T-I_2)^T \frac{1}{2}(Q+Q^T)\right)_{11},\\
m_4 &= \left((P^T-I_2) \widehat \Omega (P^T-I_2)^T\right)_{11}.
\end{align*}
\item[3.] If $|\lambda_\text{new} - \lambda_\text{old}| < 10^{-6}$, stop and return $\lambda_\text{new}$. Otherwise, set
$$
\bm{\widehat B} = \bm{\widehat B_{\lambda_\text{new}}} = (1-\lambda_\text{new})\bm{\widehat B_0} + \lambda_\text{new} \bm{\widehat B_1} = (I_2 + \lambda_\text{new} (Q - I_2)) \bm{\widehat B_0}.
$$
Set $\lambda_\text{old} \coloneqq \lambda_\text{new}$ and return to step 2.
\item[4.] The final estimate of $\bm y$ is computed by subtracting $\bm{\widehat B}$ from $\bm{\widehat y}$
\end{itemize}
Details of the derivation of the formulas in step 2 can be found in Appendix A3 in
\citetext{delden2015errors_details}. Note that we do not explicitly construct the bootstrap estimators $\bm{\widehat B_R^*}$ and $\widehat V_R^*$, but use the analytical expressions for $\bm{\widehat B_\infty^*}$ and $\widehat V_\infty^*$. For large $R$, both approaches will yield the same result, but the latter is computationally more efficient. It should also be noted that the above iterative procedure is performed for each of the years 2014, 2015 and 2016 separately. The same (estimated) matrix $P = \widehat P$ is used for each year. The optimal value of $\lambda$ might differ across years, as it depends on the annual turnover figures.

We conclude Section \ref{sec_methods_correction} by demonstrating how to estimate the standard deviation of the final estimate of $\bm{y}$. First, observe that the estimate $\bm{\widehat y} - \bm{\widehat B_\lambda}$ for the optimal value of $\lambda$ can explicitly be written as
\begin{align*}
\bm{\widehat y} - \bm{\widehat B_\lambda} &= \bm{\widehat y} - (I_2 + \lambda(Q-I_2)) \widehat B_0 \\
&= \bm{\widehat y} - (I_2 + \lambda(Q-I_2)) (P^T - I_2) \bm{\widehat y} \\
&= \left(2I_2 - P^T - (Q - I_2)(P^T-I_2)\right) \bm{\widehat y}.
\end{align*}
Then, it follows that the covariance matrix of $\bm{\widehat y} - \bm{\widehat B_\lambda}$ can be written as
$$
V(\bm{\widehat y} - \bm{\widehat B_\lambda}) = \left(2I_2 - P^T - (Q - I_2)(P^T-I_2)\right) V(\bm{\widehat y}) \left(2I_2 - P^T - (Q - I_2)(P^T-I_2)\right)^T.
$$
The covariance matrix $V(\bm{\widehat y})$ of $\bm{\widehat y}$ can be estimated by
$$
\widehat \Omega = \widehat V_\infty^*(\bm{\widehat y}) = \text{diag}\left(P^T \bm{\widehat k}\right) - P^T \text{diag}\left(\bm{\widehat k}\right) P,
$$
in which, again, $\bm{\widehat k} = \sum_i \bm{\widehat a_i} y_i^2$. Finally, the standard deviation of the first component of $\bm{\widehat y} - \bm{\widehat B_\lambda}$ will be estimated by the square root of the value in position $(1,1)$ in the $2\times 2$-matrix
$$
\widehat V(\bm{\widehat y} - \bm{\widehat B_\lambda}) = \left(2I_2 - P^T - (Q - I_2)(P^T-I_2)\right) \widehat \Omega \left(2I_2 - P^T - (Q - I_2)(P^T-I_2)\right)^T.
$$
As the values of $\bm{y_M}$ are not stochastic, this concludes the derivation of the standard deviation of the final estimate of $\bm{y}$.

\subsection{The Proposed Data-Driven Supply-Side Approach}\label{sec_methods_summary}
The proposed data-driven supply-side approach for measuring cross-border internet purchases within the EU can be summarized as follows. Based on EU VAT legislation, the starting point is a dataset of tax returns filed by foreign companies established within the EU. Tax returns are \emph{supply-side} data as they contains company sales. Then, the challenge was to identify webshops within the dataset of tax returns. We solved this problem in two phases. In the first phase, we implemented approximate string matching techniques to merge the set of tax returns to a business register of retail companies established within the EU. The merging can be viewed as \emph{data-driven} record-linkage, as we optimized the performance of the approximate string matching using machine learning algorithms. In the second phase, we used web scraping in combination with (\emph{data-driven}) machine learning to assess whether a company is a webshop. The outcomes of the two phases are combined to obtain a more accurate estimate of cross-border internet purchases. Moreover, we use the data to estimate the bias and standard deviation of the estimate. Thus, the proposed methods applied on the proposed data yield our data-driven supply-side approach for measuring cross-border internet purchases within the EU.




\section{Results}\label{sec_results}

The goal of this section is to present our main findings in applying the proposed approach to estimate cross-border internet purchases within the EU by Dutch consumers. The section is structured as follows. In Section \ref{sec_results_br}, the results of training the classification algorithms to estimate the industry class by business register are presented.  In Section \ref{sec_results_scraper}, the same is presented for estimating the industry class by web sites. In Section \ref{sec_results_final}, we present the results of estimating cross-border internet purchases by Dutch consumers. It contains the most relevant results of the paper. Finally, in Section \ref{sec_results_comparison}, we compare our results to currently available estimates of cross-border internet purchases by Dutch consumers. We interpret and discuss the differences of the estimates.


\subsection{Results from Estimating the Industry Class by Business Registers}\label{sec_results_br}

In this section, we aim to present the main findings of estimating the industry class $s_i^\text{BR}$ by business registers. Recall from Paragraph D in Subsection \ref{sec_methods_model_br} that we compared ten different machine learning algorithms (Table \ref{tab_methods_ml_algs}), each evaluated using multiple parameter settings (Table \ref{tab_methods_ml_params}). For each algorithm, we have selected the parameter settings that are optimal in predicting $s_i^\text{BR}$, based on the mean F1 score from the stratified 5-fold cross-validation. The results for Multinomial Naive Bayes are not shown, as this algorithm assumes discrete, multinomially distributed features, while the features are distances between strings and thus continuous.

The optimal parameter settings and corresponding scoring metrics for training the specified classification algorithms to estimate $s_i^\text{BR}$ are presented in Table \ref{tab_res_orbis_train}. The algorithms are ranked with respect to the optimal mean F1 score over the folds in the stratified 5-fold cross-validation. The standard deviation in scores over the five folds is shown in parentheses. Note that the differences in mean goodness of fit between the algorithms are small. Furthermore, the standard deviations in scores over the folds are small.

The algorithm we choose to use is Support Vector Classification with Radial Basis Function Kernel (RBFSVC), with parameters $C = 100$, $\gamma = 1$ and the balanced class weighing scheme. Observe that this choice not only maximizes mean F1-score, but also mean precision and mean recall. In particular, the algorithm does not falsely predict positives on the training dataset. Moreover, the local behaviour of the mean F1 score of RBFSVC, as a function of the parameters $C$ and $\gamma$, is stable around the optimal parameters. Details on the local behaviour can be found in Appendix \ref{sec_app}.

\begin{table}[t]
\centering
\caption{Mean ($\pm$ standard deviation) of scores for optimal parameter settings for each of the specified algorithms estimating $s_i^\text{BR}$. The scoring function F1 is used to optimize across the parameter settings in the parameter grid. Each parameter setting is evaluated using stratified 5-fold cross-validation. The mean F1 score is used to rank the results.}
\label{tab_res_orbis_train}
\begin{threeparttable}
\begin{tabular}{llrrrrr}
\hline
\thead{Algorithm} & \thead{Optimal Parameters} & \thead{F1} $\downarrow$ & \thead{Precision} & \thead{Recall}\\
\hline
RBFSVC & $C = 100$, $\gamma = 1$ & $\bm{0.97} \; (\pm \, 0.03)$ & $\bm{1.00} \; (\pm \, 0.00)$ & $\bm{0.94} \; (\pm \, 0.05)$ \\
GB & $n = 50$, $d = 1$, $\lambda = 0.01$ & $0.95 \; (\pm \, 0.02)$ & $0.98 \; (\pm \, 0.03)$ & $0.92 \; (\pm \, 0.03)$ \\
kNN & $k = 3$ & $0.95 \; (\pm \, 0.03)$ & $0.98 \; (\pm \, 0.03)$ & $0.93 \; (\pm \, 0.04)$ \\
LinSVC & $C = 0.01$ & $0.94 \; (\pm \, 0.02)$ & $0.97 \; (\pm \, 0.03)$ & $0.91 \; (\pm \, 0.03)$ \\
LDA & & $0.94 \; (\pm \, 0.03)$ & $\bm{1.00} \; (\pm \, 0.00)$ & $0.89 \; (\pm \, 0.05)$ \\
LR & $C = 1$, L1-penalty & $0.94 \; (\pm \, 0.03)$ & $0.97 \; (\pm \, 0.03)$ & $0.92 \; (\pm \, 0.03)$ \\
AB & $n = 100$, $d = 1$, $\lambda = 0.1$ & $0.94 \; (\pm \, 0.04)$ & $0.96 \; (\pm \, 0.04)$ & $0.93 \; (\pm \, 0.04)$ \\
RF & $n = 50$, $d = 4$ & $0.94 \; (\pm \, 0.04)$ & $0.95 \; (\pm \, 0.04)$ & $0.93 \; (\pm \, 0.04)$ \\
QDA & & $0.93 \; (\pm \, 0.02)$ & $\bm{1.00} \; (\pm \, 0.00)$ & $0.87 \; (\pm \, 0.03)$ \\
\hline
\end{tabular}
\end{threeparttable}
\end{table}


\subsection{Results from Estimating the Industry Class by Web Sites}\label{sec_results_scraper}

Similar to Section \ref{sec_results_br}, the aim of this section is to present the main findings of estimating the industry class $s_i^\text{W}$ by web sites. The optimal parameter settings and corresponding scoring metrics for training the specified algorithms to predict $s_i^\text{W}$ are presented in Table \ref{tab_res_scraper_train}. Again, the algorithms are ranked with respect to the optimal mean F1 score over the folds in the stratified 5-fold cross-validation. The standard deviation in scores over the five folds is shown in parentheses. Observe that Multinomial Naive Bayes (MNB) and Quadratic Discriminant Analysis (QDA) perform considerably less well than the other algorithms. Moreover, observe that the linear algorithms (LR, LinSVC, LDA) perform less well than the remaining algorithms. It suggests that a linear separation of the data points in higher dimensional space does not yield the best classification on unseen data. Note that the ensemble methods (AB, GB, RF) outperform the other methods. Furthermore, note that the scores for predicting $s_i^\text{W}$ are much lower than those for predicting $s_i^\text{BR}$. It should also be noted that the standard deviations in scores are relatively high. Both might be due to the fact that in many cases, the URLfinding software does not return a URL that is likely to correspond to the input legal company name.

The algorithm we choose to use is Random Forest (RF), with parameters $n = 200$, $d=1$ and the balanced class weighing scheme. The reason for this choice is that RF maximizes mean precision. Moreover, the local behaviour of the F1 score of RF, as a function of the algorithm parameters, is more stable around the optimal parameters compared to the local behaviour for AB and GB. Details can be found in Appendix \ref{sec_app}.

\begin{table}[t]
\caption{Mean ($\pm$ standard deviation) of scores for optimal parameter settings for each of the specified algorithms predicting $s_i^\text{W}$. The scoring function F1 is used to optimize across the parameter settings in the parameter grid. Each parameter setting is evaluated using stratified 5-fold cross-validation. The mean F1 score is used to order the optimal algorithms.}
\label{tab_res_scraper_train}
\begin{threeparttable}
\begin{tabular}{llrrr}
\hline
\thead{Algorithm} & \thead{Optimal Parameters} & \thead{F1} $\downarrow$ & \thead{Precision} & \thead{Recall}\\
\hline
AB & $n = 100$, $d = 1$, $\lambda = 0.1$, bal. & $\bm{0.80} \; (\pm \, 0.11)$ & $0.82 \; (\pm \, 0.10)$ & $0.78 \; (\pm \, 0.12)$ \\
GB & $n = 200$, $d = 1$, $\lambda = 0.1$ & $0.79 \; (\pm \, 0.10)$ & $0.80 \; (\pm \, 0.09)$ & $0.78 \; (\pm \, 0.12)$ \\
RF & $n = 200$, $d = 1$, bal. & $0.78 \; (\pm \, 0.10)$ & $\bm{0.85} \; (\pm \, 0.14)$ & $0.76 \; (\pm \, 0.16)$ \\
kNN & $k = 35$ & $0.76 \; (\pm \, 0.09)$ & $0.81 \; (\pm \, 0.04)$ & $0.73 \; (\pm \, 0.17)$ \\
RBFSVC & $C = 1$, $\gamma = 0.1$, bal. & $0.76 \; (\pm \, 0.11)$ & $0.78 \; (\pm \, 0.08)$ & $0.76 \; (\pm \, 0.18)$ \\
LR & $C = 1$, L1-penalty & $0.75 \; (\pm \, 0.12)$ & $0.76 \; (\pm \, 0.10)$ & $0.76 \; (\pm \, 0.18)$ \\
LinSVC & $C = 0.01$ & $0.74 \; (\pm \, 0.10)$ & $0.77 \; (\pm \, 0.07)$ & $0.73 \; (\pm \, 0.17)$ \\
LDA &  & $0.74 \; (\pm \, 0.13)$ & $0.69 \; (\pm \, 0.14)$ & $\bm{0.81} \; (\pm \, 0.17)$ \\
MNB & $\alpha = 10^{-10}$ & $0.70 \; (\pm \, 0.12)$ & $0.71 \; (\pm \, 0.15)$ & $0.71 \; (\pm \, 0.12)$ \\
QDA &  & $0.67 \; (\pm \, 0.15)$ & $0.63 \; (\pm \, 0.12)$ & $0.73 \; (\pm \, 0.21)$ \\
\hline
\end{tabular}
\end{threeparttable}
\end{table}


\subsection{Estimating Cross-Border Internet Purchases}\label{sec_results_final}

This section contains the most relevant results of the paper, as it presents, in Table \ref{tab_results_final}, the final estimates of cross-border internet purchases within the EU by Dutch consumers.

Recall that to obtain the results of this section, the algorithm chosen in Section \ref{sec_results_br} was (re)trained on the entire training dataset indexed by $I_M$. It resulted in a model $\widehat s_i^\text{BR}$ that was qualified to predict $s_i$ on the remaining part of the dataset of tax returns, indexed by $I \backslash I_M$. Similarly, a model $\widehat s_i^\text{W}$ was trained using the algorithm chosen in Section \ref{sec_results_scraper}. The two models were combined into a final model $\widehat s_i$ as described in Subsection \ref{sec_methods_model_final}. The comparison between the model $\widehat s_i$ and the true, observed values $s_i$ on the test dataset, yields the values TP~=~8, FP~=~4, TN~=~62 and FN~=~5. It follows that
$$
\widehat P = 
\begin{pmatrix}
8/13 & \;\; 5/13\\
4/66 & 62/66
\end{pmatrix}
\approx
\begin{pmatrix}
0.615 & 0.385 \\
0.061 & 0.939
\end{pmatrix}.
$$
The main results of the paper are shown in Table \ref{tab_results_final}. The values $y_M$ contain the total cross-border internet purchases at companies in the set $I_M$. The categorization for companies in $I_M$ has been manually determined and can be considered free from errors. The values $\widehat y$ contain the additional estimated cross-border internet purchases at companies in the set $I \backslash I_M$. The values $\lambda_\text{opt}$ contain the optimal values of $\lambda$ in minimizing the mean squared error of the estimated bias of $\widehat y$. Note that all optimal values of $\lambda$ are equal to 0, meaning that the increased variance dominates the decreased squared bias of $\bm{\widehat B_1}$ compared to $\bm{\widehat B_0}$. This is due to the relatively high off-diagonal values in the matrix $\widehat P$. The values $\widehat B_{\lambda_\text{opt}}$ represent the estimated bias of $\widehat y$ for the optimal value $\lambda = \lambda_\text{opt}$. Note that the bias strongly differs across the three years. The values $y$ show the final estimate for the total cross-border internet purchases, computed as $y = y_M + (\widehat y - \widehat B_{\lambda_\text{opt}})$. The final column in Table \ref{tab_results_final} contains the standard deviation of $y$, estimated as outlined at the end of Section \ref{sec_methods_correction}.

In the Netherlands, total household consumption on retail goods (food and durable goods, codes 1000 up until and including 3000) in 2016 is equal to EUR~87,206 million, according to Statistics Netherlands (\url{https://opendata.cbs.nl}). 
Statistics Netherlands does not publish the total online consumption of goods by Dutch consumers. The only currently available estimate is by Thuiswinkel.org and GfK and it is based on consumer surveys. The estimate for 2016 equals EUR 11.01 billion. It seems possible that just over 12 per cent of online consumption by Dutch consumers is spent at foreign webshops established within the EU. Besides, Statistics Netherlands does publish year-on-year growth figures on online retail sales by Dutch webshops. In 2016, this year-on-year growth was equal to 22.1 per cent. 
It is quite similar to the growth of 21.2 per cent that we find by comparing the values of $y$ in 2015 and 2016 as presented in Table \ref{tab_results_final}.

Reflecting on our main findings, we note that the standard deviation of the final estimate would still be too large for official statistical purposes. However, as discussed more thoroughly in Section \ref{sec_results_comparison}, our main findings prove to be a significant improvement compared to currently available alternative estimates.

\begin{table}[t]
\centering
\caption{Final results in millions of euros. Inconsistency in computing $y$ from $y_M$, $\widehat y$ and $\widehat B_{\lambda_\text{opt}}$ is due to rounding.  }
\label{tab_results_final}
\begin{threeparttable}
\begin{tabular}{lrrrrrr}
\hline
Year & $y_M$ & $\widehat y$ & $\lambda_\text{opt}$ & $\widehat B_{\lambda_\text{opt}}$ & $y$ & $\text{Std}(y)$\\
\hline
2014 & 405 & 495 & 0 & 63 & 837 & 97 \\
2015 & 565 & 586 & 0 & 21 & 1,132 & 101 \\
2016 & 725 & 667 & 0 & 19 & 1,372 & 110 \\
\hline
\end{tabular}
\end{threeparttable}
\end{table}

\subsection{Comparison with Demand-Side Approach}\label{sec_results_comparison}

In Section \ref{sec_introduction}, we have claimed that a our data-driven supply-side approach would be more accurate than demand-side approaches to estimate cross-border internet purchases within the EU. To support this claim, we compare our results for the Netherlands to the results of the consumer-survey approach by market research institute GfK (commissioned by Thuiswinkel.org on behalf of Ecommerce Europe). In 2016, total cross-border online consumption by Dutch consumers according to GfK was equal to EUR 637 million, EUR 190 million of which was spent in China and EUR 70 million in the United States. This implies that at most EUR 377 million was spent within the EU. Note that this figure includes online consumption of both goods and services.

Moveover, the fraction of online consumption of goods in the total online consumption in 2016, as reported by GfK, was EUR 11.01 billion / EUR 20.16 billion = 0.55. We assume that this proportion is independent of the location of the purchased goods or services. As a result, cross-border online purchases of goods within the EU, according to GfK, would approximately equal EUR 206 million in 2016.

We, however, find EUR 1,372 million for 2016 with a standard deviation of EUR 110 million. The estimate is more that 6 times as high as that of GfK. Our results thus show the severe downward bias in using demand-side approaches to estimate cross-border online consumption and it motivates the implementation of our approach in other EU member states.




\section{Main Conclusion and Future Work}\label{sec_conclusion}
We have proposed to use tax returns, business registers and internet data to measure cross-border internet purchases within the EU. We have implemented data-driven methods to combine these supply-side data in a computationally efficient manner. The empirical results show that the proposed approach leads to a strong improvement of existing approaches based on consumer surveys, as it overcomes the language bias as described in Section \ref{sec_introduction}. In particular, we find that cross-border internet purchases by Dutch consumers within the EU in 2016 equal EUR 1,372 million. Market research institute GfK (commissioned by Thuiswinkel.org on behalf of Ecommerce Europe) estimate it to be approximately EUR 206 million based on consumer surveys. The language bias in consumer surveys thus results in a downward bias of a factor of more than 6 in cross-border internet purchases.

The proposed approach only requires foreign companies' tax returns to contain the legal company name and the turnover taxed at high or low tariff. Due to EU VAT legislation these data are available in every EU member state. Note that we do not require that the economic activity of a company is accurately available in filed tax returns. The training and test set could even be constructed without any known economic activity, by viewing all companies as belonging to the same class and following the construction as described in Sections \ref{sec_data_train} and \ref{sec_data_test}. Moreover, we do not assume that the URL of the home page of a company is available in filed tax returns. Furthermore, the additional data (the BR and internet data) required by the proposed approach are obtained from open data sources. Hence, here we arrive at our main conclusion, viz. that the proposed approach is expected to be applicable in any other EU member state and thus can estimate cross-border internet purchases within the EU.

Further applications of the proposed supply-side approach include revealing the structure of the cross-border online retail market in any EU member state. Our approach directly returns a list of foreign webshops and their annual cross-border internet sales to the observing country. If the information on domestic webshops active on the country's online market is complemented, it can be used to analyse the market structure. A related research topic is the export of all webshops established in a single EU member state. It is related as it is the supply-side counterpart of the cross-border online consumption market within a country. It might be interesting to compare the two market structures within individual countries and compare the market structures between countries within the EU.

Future work on measuring cross-border internet purchases within the EU might focus on improving the website-based predictions of company classifications. The empirical results show that it is the weakest part of the proposed approach, as the F1 scores for website-based predictions are lower than the F1-scores for the BR-based predictions. The main improvement could be achieved by sophisticating the URLfinding algorithm in order to increase the recall of the predictions. We suggest to enlarge the training set of the URLfinding algorithm from Dutch to European websites using the company names and URLs registered in the BR. We consider this improvement outside the scope of the current paper, as the results of our supply-side approach already show a strong improvement compared to existing consumer-survey approaches.

To summarize, we have proposed a data-driven supply-side approach for measuring cross-border internet purchases within the EU. Our main findings motivate the implementation of our approach in other EU member states. As our approach is based on EU VAT legislation, international comparability will be guaranteed. Ultimately, it will lead to more accurate estimates of cross-border internet purchases within the entire EU.




\section*{Acknowledgements}
We would like to express our sincere gratitude to Arjan van Loon for many fruitful discussions on data analysis and research methods. Moreover, we are grateful to Dick Windmeijer for developing the URLfinding software and running it on the data. We also thank Guido van den Heuvel for developing the web scraping software and running it on the data. Furthermore, we thank Myra Bissumbhar and Marja Groen for manually categorizing companies in the training and test datasets. Finally, we are grateful to Willem de Jong and Jos Erkens for helpful discussions on EU VAT legislation. The research was funded by Statistics Netherlands and the Dutch Ministry of Economic Affairs and Climate Policy.




\bibliographystyle{oxford}

\bibliography{cross-border}

\newpage



\appendix

\section*{Appendix}

\section{Local Behaviour around Optimal Parameters}\label{sec_app}

This appendix contains additional results on the local behaviour of the mean and standard deviation of F1 scores obtained from the $5$-fold cross-validation around the optimal parameters for the optimal algorithm. The results for $\widehat s_i^\text{BR}$ and $\widehat s_i^\text{W}$ are presented separately.

\begin{figure}[h]
\centering
\subfloat[]{\includegraphics[width=0.48\textwidth]{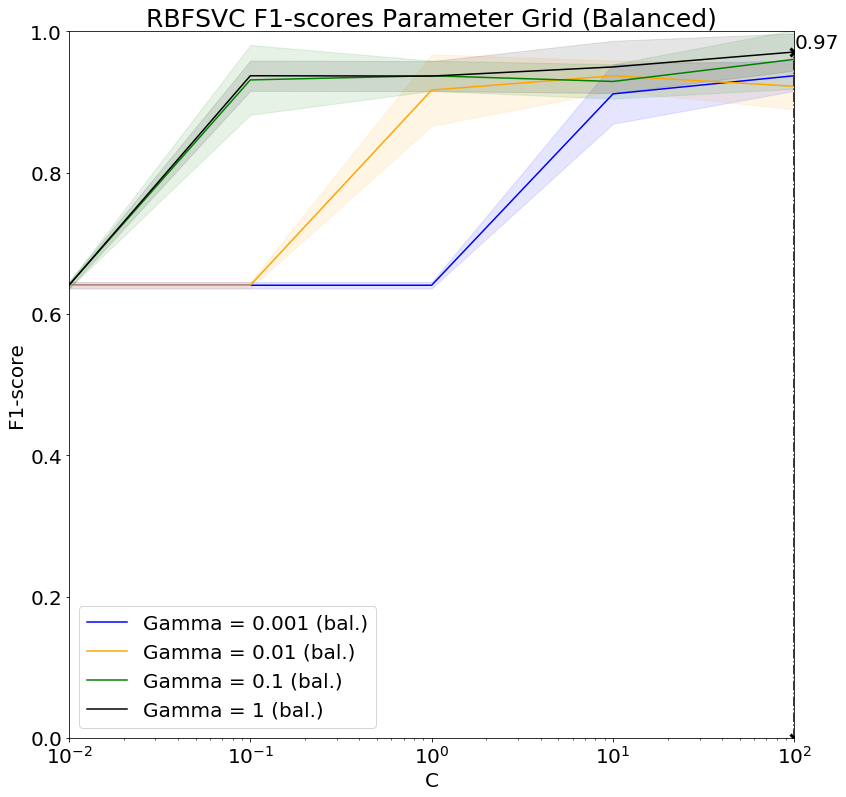}}
\subfloat[]{\includegraphics[width=0.48\textwidth]{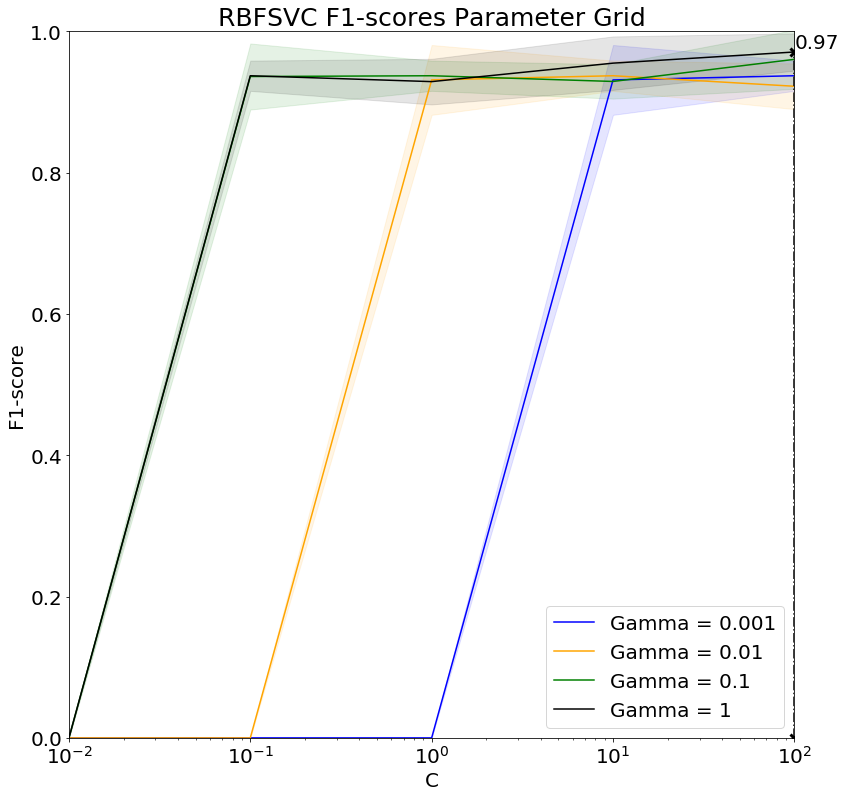}}
\caption{Mean (and standard deviation of) F1 scores for the RBFSVM algorithm trained to predict $\widehat s_i^\text{BR}$. Panel (a) displays the results for the balanced class weighing scheme and panel (b) for the uniform class weighing scheme. For comparison, the optimal parameter setting $(C = 100, \gamma = 1, \text{bal.})$ with corresponding F1 score $0.97$ is included in the figure (marked by the black cross mark at $C = 10^2$).}
\label{fig_res_orbis_RBFSVM_balanced}
\end{figure}

Note that in Fig. \ref{fig_res_orbis_RBFSVM_balanced}, the results for $C \geq 10$ hardly depend on the chosen class weighing scheme. Moreover, different choices of $\gamma$ and different choices of $C$, given $C \geq 10$, only minimally affect the mean F1 score over the 5 folds. The variance seems similar in each of these parameter settings as well. Thus, the mean F1 score is stable around the optimal parameter setting.

\begin{figure}[h!]
\centering
\subfloat[]{\includegraphics[width=0.45\textwidth]{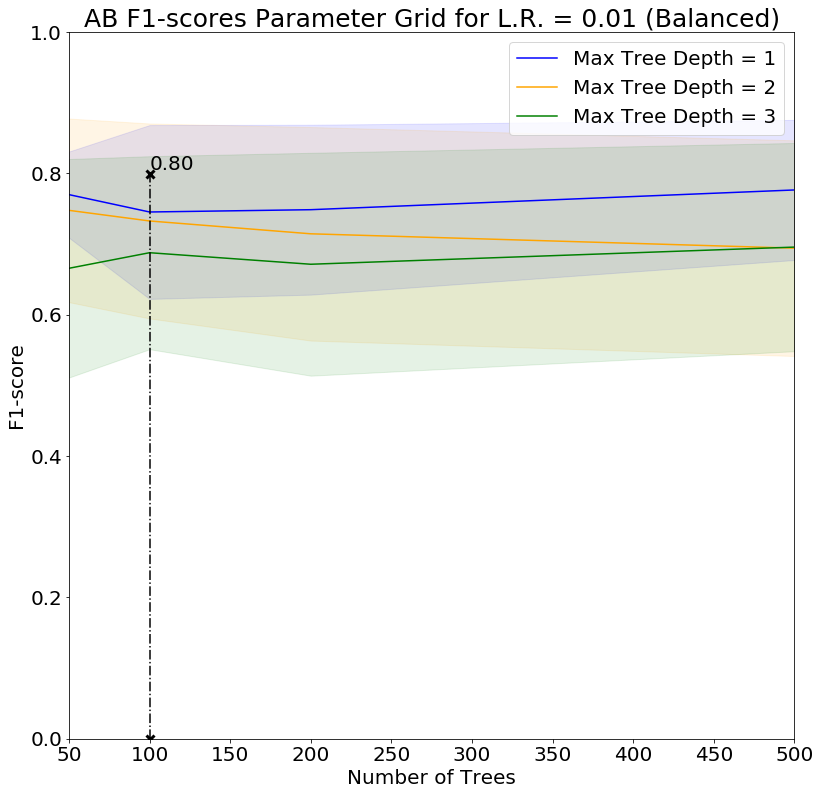}}
\subfloat[]{\includegraphics[width=0.45\textwidth]{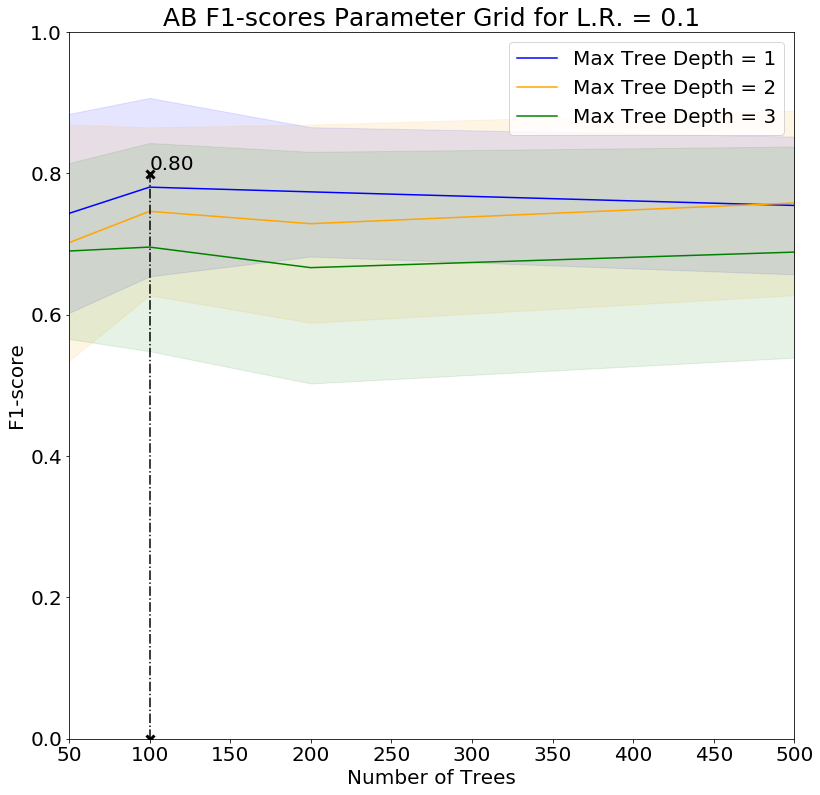}}\\
\subfloat[]{\includegraphics[width=0.45\textwidth]{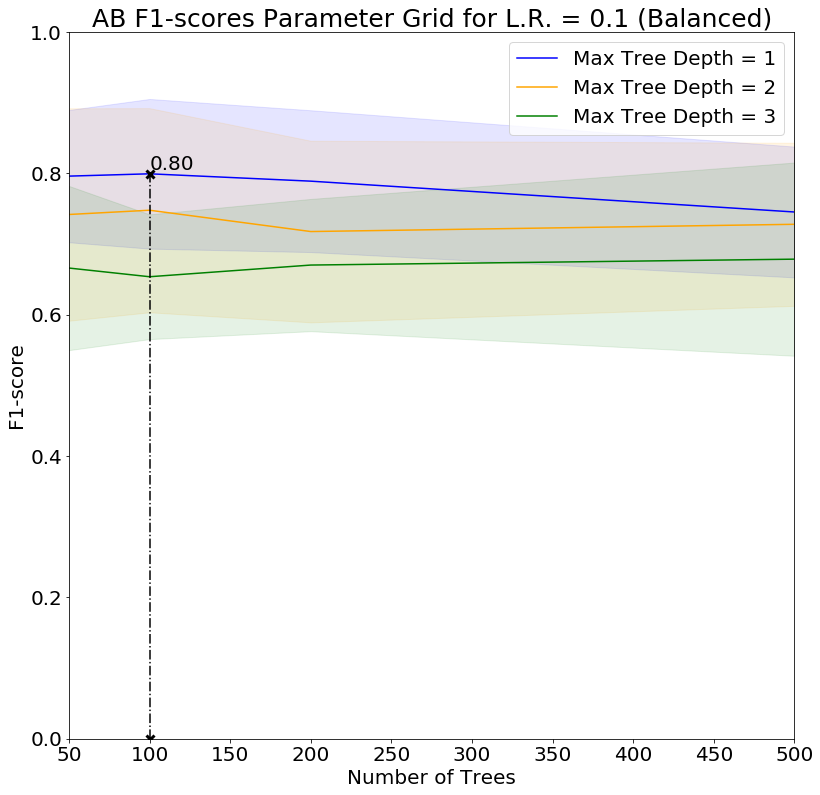}}
\subfloat[]{\includegraphics[width=0.45\textwidth]{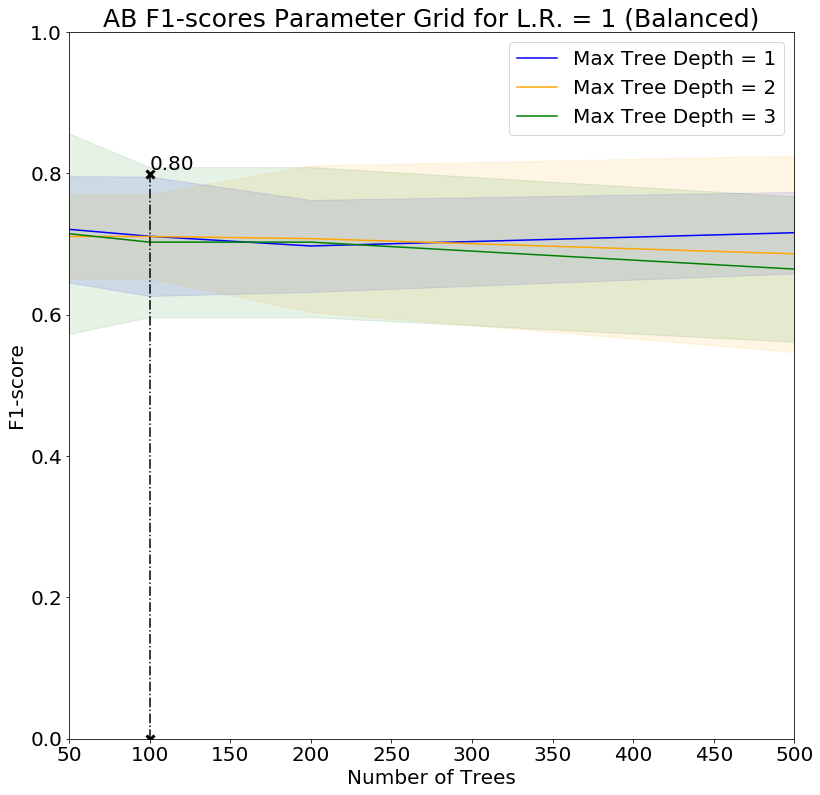}}
\caption{Mean F1 scores for AB trained to predict $\widehat s_i^\text{W}$. Panels (a), (c) and (d) contain the results for the specified parameter grid with the learning rate parameter $\lambda$ set to $0.01$, $0.1$ and $1$, respectively. Panel (b) differs from panel (c) in the value of class weighing scheme parameter. The $x$-axis depicts the number $n$ of boosted trees.}
\label{fig_res_scraper_AB}
\end{figure}

Next, we examine the local behaviour of the mean and standard deviation of F1 scores obtained from the $5$-fold cross-validation around the optimal parameters for the algorithms GB, AB and RF predicting $\widehat s_i^\text{W}$. From Fig. \ref{fig_res_scraper_AB} it can be derived that increasing the maximum tree depth $d$ from the optimal value $d = 1$ negatively impacts the goodness-of-fit as measured by F1. Moreover, in panels (a) and (d) the mean F1 score is much lower for $d = 1$ compared to panel (c). Thus, the results from AB are not stable around the optimal maximum tree depth $d = 1$ and not around the optimal learning rate $\lambda = 0.1.$ The results do seem independent of the choice of the class weighing scheme, when comparing panels (b) and (c).

\begin{figure}[ht]
\centering
\subfloat[]{\includegraphics[width=0.45\textwidth]{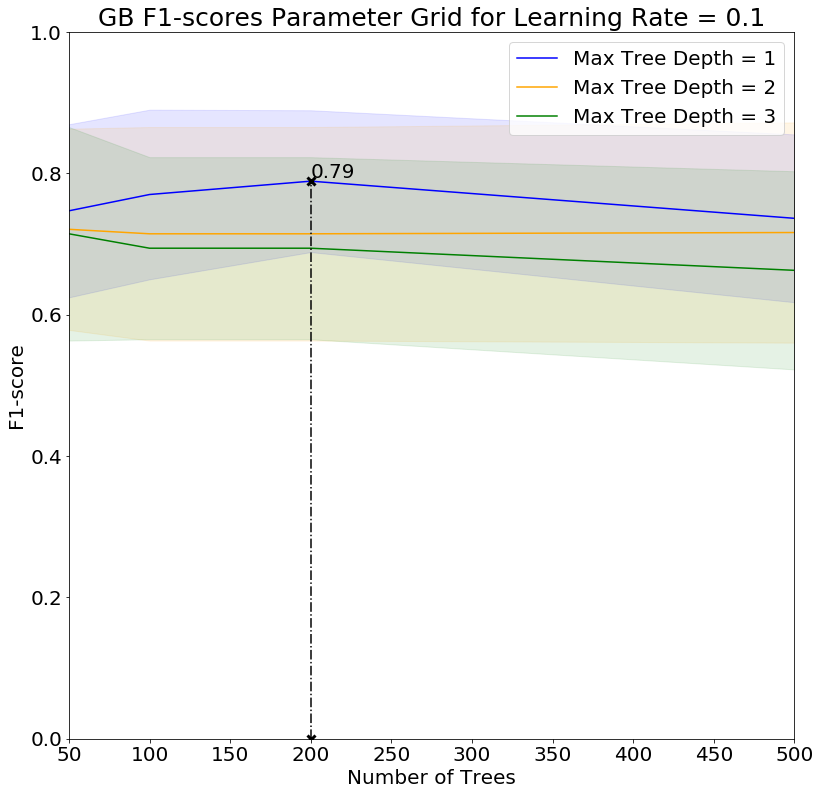}}
\subfloat[]{\includegraphics[width=0.45\textwidth]{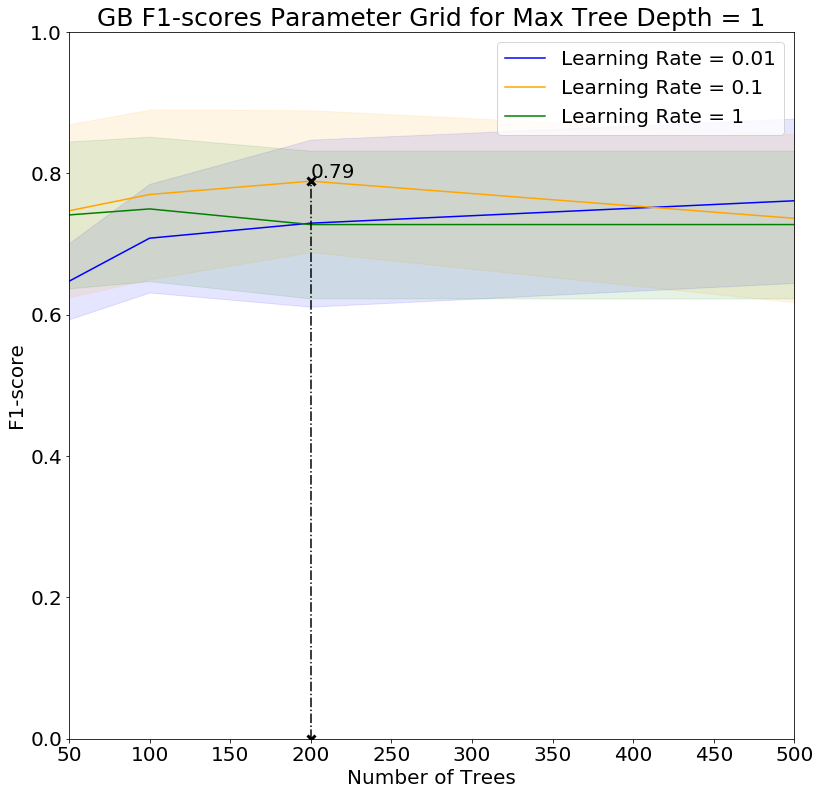}}
\caption{Mean F1 scores for GB trained to predict $\widehat s_i^\text{W}$. Panel (a) contains the results for different values of $d$ and the fixed optimal value of $\lambda$. Panel (b) displays the results for different values of $\lambda$ and the fixed optimal value of $d$. The $x$-axis depicts the number $n$ of boosted trees.}
\label{fig_res_scraper_GB}
\end{figure}

From Fig. \ref{fig_res_scraper_GB}, we may conclude that the mean F1 scores of GB are not very stable around the optimal parameter setting $(n = 200, d = 1, \lambda = 0.1)$. In panel (a), it can be seen that increasing the maximum depth $d$ from the optimal value $d = 1$ while fixing the optimal value $\lambda = 0.1$ of $\lambda$, leads to a drop in mean F1 score. The same holds for changing the optimal value $\lambda  = 0.1$ while fixing $d = 1$.

\begin{figure}[ht]
\centering
\subfloat[]{\includegraphics[width=0.45\textwidth]{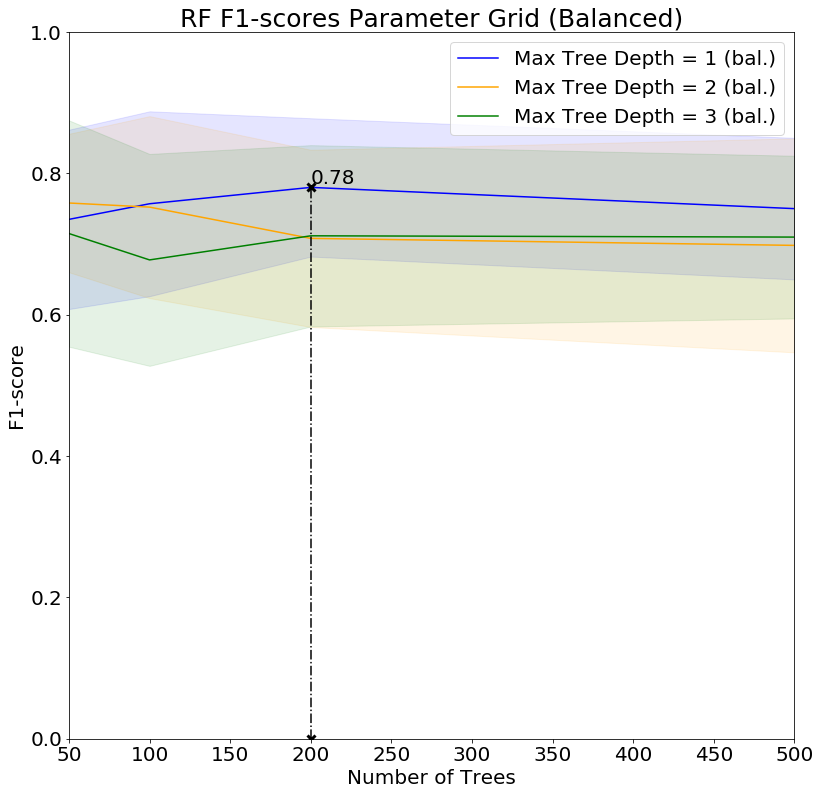}}
\subfloat[]{\includegraphics[width=0.45\textwidth]{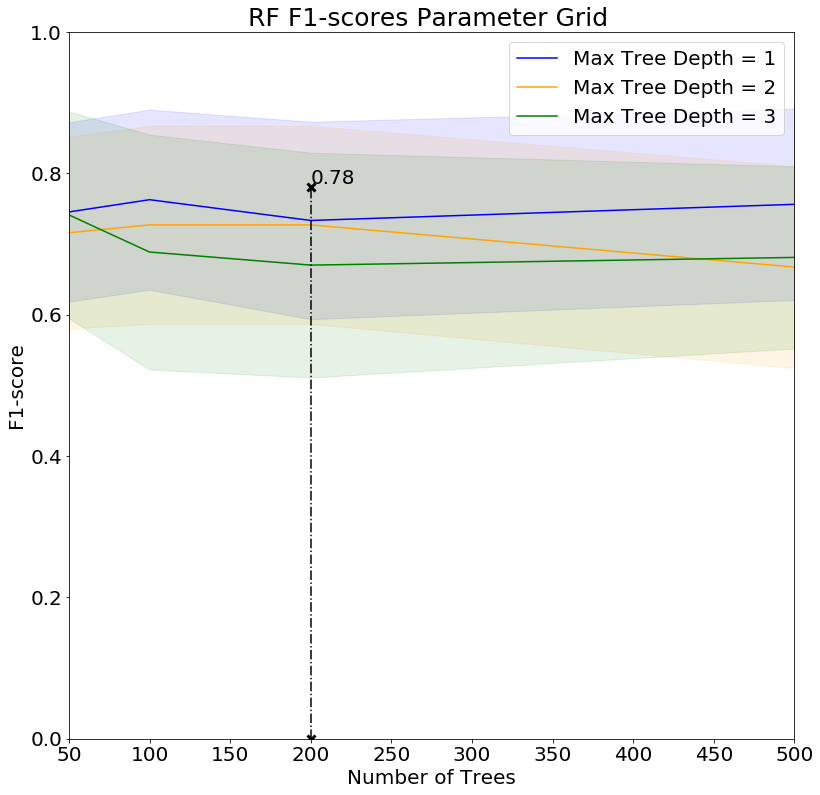}}
\caption{Mean F1 scores for RF trained to predict $\widehat s_i^\text{W}$. Panel (a) and (b) contain the results for the balanced and uniform class weighing scheme, respectively. The $x$-axis depicts the number $n$ of trees in the forest.}
\label{fig_res_scraper_RF}
\end{figure}

Finally, we present the results of RF on the training dataset in Fig. \ref{fig_res_scraper_RF}. For the balanced class weighing scheme, presented in panel (a), the results seem stable as $n$ increases. Moreover, the variance is smaller than in the uniform class weighing scheme, as displayed in panel (b). However, increasing the maximum depth $d$ from the optimal value $d = 1$ leads to a drop in mean F1 score.

\end{document}